\documentclass[5p]{elsarticle}
\makeatletter
\def\ps@pprintTitle{%
  \let\@oddhead\@empty
  \let\@evenhead\@empty
  \let\@oddfoot\@empty
  \let\@evenfoot\@empty}
\makeatother

\usepackage[utf8]{inputenc}
\usepackage[most]{tcolorbox}
\usepackage{lipsum}

\usepackage{adjustbox}    % per scalare in larghezza

\usepackage[T1]{fontenc}
\usepackage{graphicx}
\usepackage{mdframed}
\usepackage{pgfplots}
\usepackage{amsfonts}
\pgfplotsset{compat=1.18} % versione aggiornata
\usepackage{booktabs}   % per \toprule, \midrule, \bottomrule
\usepackage{textcomp}   % per \textsubscript (F\textsubscript{1})
\usepackage{siunitx}    % per \num{5039} e formattazione uniforme dei numeri
\usepackage{multirow}
\usepackage{pgfplotstable}
\usepackage{subcaption}
\pgfplotsset{compat=newest}
\usepackage{amsmath}
\usepackage{listings}
\usepackage{url}
\usepackage{caption}
\usepackage{booktabs}
\usepackage{colortbl}
\usepackage{cuted}        % \begin{strip}  … \end{strip}  (spans both cols)

\usepackage{tabularx}

\newcommand{\RQ}[1]{\tcbox[on line,boxsep=0pt,left=1pt,right=1pt,top=1pt,bottom=1pt]{#1}}

\definecolor{LightGreen}{HTML}{DFF0D8}
\definecolor{LightRed}{HTML}{F2DEDE}

\newtcolorbox{questionbox}{colback=red!15,  width=\textwidth}
\newtcolorbox{answerbox}  {colback=green!15,width=\textwidth}

\lstdefinelanguage{PromptXML}{
  morekeywords={<reasoning>,</reasoning>,<answer>,</answer>},
  sensitive=false,
  morestring=[b]",
  morecomment=[s]{<!--}{-->}
}

\lstdefinestyle{promptstyle}{
  language       = PromptXML,
  basicstyle     = \scriptsize\ttfamily,
  keywordstyle   = \color{blue!70!black}\bfseries,
  commentstyle   = \color{gray},
  stringstyle    = \color{orange!60!black},
  breaklines     = true,
  columns        = fullflexible,
  showstringspaces=false,
  frame          = single,
  rulecolor      = \color{black!50},
  framerule      = 0.5pt,
  framesep       = 6pt,
  backgroundcolor= \color{gray!5},
  captionpos     = b,          % <— caption sotto
  xleftmargin    = 0pt,
  xrightmargin   = 0pt
}

\renewcommand{\arraystretch}{1.1}  % migliora la leggibilità verticale

\journal{ }

\begin{document}

\begin{frontmatter}

    %% Title
    % \title{On Device Derivation of IoT Usage Control Policies\\ from Natural Language to U-XACML: Automating Usage Control Policy Generation with LLMs in Smart Home Environments\tnoteref{t1}}
    \title{Improving LLM Reasoning for Vulnerability Detection via \\ Group Relative Policy Optimization}
    % \tnotetext[t1]{This work has been partially supported by the Horizon Europe project MEDIATE (grant agreement 101168465), by the European Union -- NextGenerationEU within the framework of PNRR Mission 4 -- Component 2 -- Investment 1.1 under the Italian Ministry of University and Research (MUR) programme ``PRIN 2022 PNRR'' -- grant number 2022598LMZ -- AsCoT-SCE  (Assessing Compliance of IoT API for Security Critical Environments) -- CUP: H53D23003430006 and grant number P2022WEAH7 -- ASSISTANTS (Assessing Software Security, prIvacy, and SusTAiNability through Testing techniqueS) -- CUP: B53D23026270001.} 

    % %% Authors

    \author[1,3]{Marco Simoni\corref{cor1}\fnref{equal}}
    \ead{marco.simoni@iit.cnr.it}
    \author[1,2]{Aleksandar Fontana\fnref{equal}}
    \ead{aleksandar.fontana@santannapisa.it}
    \author[2]{Giulio Rossolini}
    \ead{giulio.rossolini@santannapisa.it}

    \author[2]{Andrea Saracino}
    \ead{andrea.saracino@santannapisa.it}
    \cortext[cor1]{Corresponding author}
    \fntext[equal]{These authors contributed equally to this work.}
    %% Affiliations
    \affiliation[1]{organization={Institute of Informatics and Telematics, National Research Council of Italy},
    addressline={Via G. Moruzzi 1},
    city={Pisa},
    postcode={56124},
    country={Italy}}
    
    \affiliation[2]{organization={Department of Excellence in Robotics and AI, TeCIP, Scuola Superiore Sant'Anna},
    addressline={Piazza Martiri della Libertà 33},
    postcode={56127},
    city={Pisa},
    country={Italy}}

    \affiliation[3]{organization={National Doctorate on Artificial Intelligence, Sapienza Università di Roma},
    addressline={Piazza Aldo Moro 5},
    postcode={00185},
    city={Roma},
    country={Italy}}

    \begin{abstract}
    % Improving and understanding the training dynamics of Large Language Models (LLMs) is essential to ensure their reliable behavior in safety and security-critical applications such as software vulnerability detection. Although LLMs have demonstrated impressive zero-shot capabilities in many tasks, their performance in vulnerability detection remains unstable and poses unique challenges. In particular, we analyze the problem of known LLMs over-predicting vulnerabilities while missing real ones under domain shifts. To address this challenge, this work investigates the use of a recent policy-based optimization for fine-tuning, specifically Group Relative Policy Optimization (GRPO), to shape LLM behavior through structured rule-based rewards in the field of vulnerability detection. To do so, we first adapt GRPO to this specific setting by redefining advantages and rewards from prominent datasets in this field, and evaluate its effectiveness across datasets with varying characteristics: BigVul, DiverseVul, and CleanVul. 
    
    % An extensive experimental section has been conducted to address different research questions that study the impact of using this approach in terms of generalization, reasoning improvements, and also benefits with respect to classic supervised fine-tuning (SFT). The results provide important insights for the future use of LLMs for the task of vulnerability detection.

    Improving and understanding the training dynamics and reasoning of Large Language Models (LLMs) has become essential for their deployment in AI-based security tools, such as software vulnerability detection. In this work, we present an extensive study aimed at advancing recent RL-based finetuning techniques for LLMs in the context of vulnerability detection.
    We start by highlighting key limitations of commonly adopted LLMs, such as their tendency to over-predict certain types of vulnerabilities while failing to detect others. To address this challenge, we explore the use of Group Relative Policy Optimization (GRPO), a recent policy-gradient method, for guiding LLM behavior through structured, rule-based rewards. We enable its application to the vulnerability detection task by redefining its advantage functions and reward signals using annotations from widely used datasets in the field, including \textit{BigVul}, \textit{DiverseVul}, and \textit{CleanVul}.
    The proposed methodology enables an extensive set of experiments, addressing multiple research questions regarding the impact of GRPO on generalization, reasoning capabilities, and performance improvements over standard supervised finetuning (SFT). Our findings offer valuable insights into the potential of RL-based training to enhance both the performance and reasoning abilities of LLMs in the context of software vulnerability detection.

    \end{abstract}

    %% Keywords
    
\end{frontmatter}

% \begin{figure}[t]
%   \includegraphics[width=0.49\textwidth]{images/dummy-initial-image.png}
% <  \caption{This is an idea of the image that we would like to add at the start of the paper, if makes clear that GRPO is better than everything else}
%   \label{fig:initial-image}
% \end{figure}

% Nowadays, Large Language Models (LLMs) are increasingly being adopted across a wide range of domains, where, if appropriately trained and configured, they can overcome the performance of other tools and techniques currently in use [CITE]. Among these domains, safety and security of critical scenarios, where erroneous outputs or misleading explanations can lead to undesirable system behaviors [CITE]. In particular, LLMs have also shown promising capabilities in supporting software engineering tasks such as vulnerability detection and code inspection [CITE], where both accuracy and interpretability are essential. Recent studies have demonstrated encouraging results in this direction [CITE], and several datasets have been introduced to benchmark LLM performance across diverse environments [CITE], although many remain limited in scope. In particular... DETAIL LIMITATIONS OF CURRENT APPROACHES

\section{Introduction}
Nowadays, Large Language Models (LLMs) are increasingly being adopted across a wide range of domains, where, if properly trained and configured, they can outperform traditional tools and techniques~\cite{ai_tools}. 
%Among the key challenges in recent LLM research, enhancing and understanding LLM reasoning is crucial, as it enables models to produce more verifiable and explainable responses, an essential requirement for their deployment in safe and secure application scenarios~\cite{ghazal2024cybersecurity, yuan2024back}.
Within the growing interest in AI for security, LLMs have been explored for supporting software engineering tasks such as vulnerability detection and code inspection~\cite{zhou2025large}. In these scenarios, both detection accuracy and explainability of reasoning are crucial to assist developers and stakeholders in identifying potential attacks or implementation errors. Recent studies have shown encouraging results in this direction~\cite{graph_vul_det, zhou2025large}, demonstrating that off-the-shelf LLMs can understand certain vulnerability patterns and provide preliminary reasoning about their nature. 
%In parallel, multiple datasets have been introduced to benchmark the performance of LLMs across diverse environments and code types~\cite{bigvul, cleanvul, diversevul}.

However, despite these preliminary yet promising achievements of LLMs in this context, two critical aspects require further exploration: (i) the ability of an LLM to generalize to new code domains (e.g., previously unseen vulnerabilities not covered during training); and (ii) its capacity to provide clear and human-understandable reasoning for its predictions.
For instance, prior work~\cite{Analysing_LLM_Capabilities} has shown that LLMs often fail to detect subtle memory-related flaws, especially when such vulnerabilities are obfuscated by project-specific coding styles.
These challenges are even more pronounced in small and medium-sized LLMs, which, although limited in capacity, are increasingly of interest for industrial and private use due to their lower computational cost and resource requirements~\cite{zhou2025large}. 

To tackle these challenges, several strategies have been proposed. Supervised finetuning (SFT), for example, has been shown to improve recall~\cite{fine-tuninng-llm}, but often at the cost of overfitting and limited reasoning generalization to novel or unfamiliar codebases. Other directions have explored the use of Retrieval-Augmented Generation (RAG) methods~\cite{rag-vul-dete}, which enhance predictions by incorporating external knowledge. However, they may introduce latency, require carefully engineered setups, and are highly dependent on the quality and coverage of the retrieval index.

Despite the efforts posed by the above mentioned strategies, alternative directions need to be explored, such as the use of reinforcement learning (RL) to shape LLM behavior, which has shown impressive improvements in mathematical reasoning as remarked by the advent of recent models like DeepSeek~\cite{guo2025deepseek, shao2024deepseekmath}.
In the context of vulnerability detection, RL-based finetuning remain largely unexplored, as they requires carefully designing reward functions and policy update mechanisms that align with security-relevant objectives, particularly in a domain where training data is often sparse, noisy, or of low quality~\cite{NEURIPS2024_1e38b2a0}.

\medskip
\noindent \paragraph{This work} Building on these observations, this work extends a previous analysis~\cite{FontanaSimoni2025} and, {guided by three key research questions, aims to address current gaps in how recent LLMs are trained and how effectively they generalize and reason in the context of vulnerability detection.

First, we ask: \RQ{RQ1} \textit{Can a small instruction tuned LLM reason effectively about software vulnerabilities without additional finetuning?} Addressing this question allows us to assess the zero-shot capabilities of lightweight models on complex reasoning tasks.

Second, we explore: \RQ{RQ2} \textit{Can we train the model to use its own reasoning to identify vulnerabilities?} This allows us to investigate whether structured, self-generated reasoning can improve detection performance.}
We tackle such challenge by incorporating self-reasoning during LLM training, enabling the use of Group Relative Policy Optimization (GRPO)~\cite{shao2024deepseekmath}, a reinforcement learning strategy designed to guide model behavior through structured, rule-based rewards. 
To this end, we design a \textit{modular reward function} that scores model outputs along three dimensions: \textit{Formatting}, \textit{Correctness}, and \textit{Reasoning}. These scores are combined using a dynamic weighting scheme designed to mitigate issues such as \textit{reward hacking}~\cite{wang2025beyond,reward_hacking}.
% \textcolor{blue}{To this end, we design a \textit{modular reward function} that scores model outputs along three dimensions: \textit{Formatting}, to ensure structural correctness; \textit{Correctness}, to verify the final answer; and \textit{Reasoning}, to evaluate the logical consistency of intermediate steps. These scores are combined using a time-dependent weighting that initially prioritizes formatting, then gradually shifts focus toward correctness and reasoning. This dynamic scheme mitigates issues such as \textit{reward hacking}~\cite{wang2025beyond,reward_hacking}.}

Finally, we consider: \RQ{RQ3} \textit{In what ways does GRPO differ from (and potentially improve upon) traditional supervised finetuning for code vulnerability detection?}
We validate the proposed approach through an extensive suite of experiments aimed at evaluating the benefits of GRPO in terms of generalization, robustness, and alignment with vulnerability-specific reasoning, demonstrating significant improvements over traditional SFT.

All experiments and analyses are conducted using three instruction-tuned Small Language Models (SMLs) (LLaMA 8B~\cite{llama3}, LLaMA 3B~\cite{llama3}, and Qwen 2.5 3B~\cite{yang2025qwen3}) evaluated on three widely used vulnerability detection datasets: \textsc{BigVul}~\cite{bigvul}, \textsc{DiverseVul}~\cite{diversevul}, and \textsc{CleanVul}~\cite{cleanvul}.

% \textcolor{blue}{Our experiments show that GRPO with our proposed modular reward function significantly outperforms the zero-shot baseline: recall of the vulnerable class increases from 0.21 → 0.56 for \textit{LLaMA 3B}/\textit{BigVul}, accuracy from 68.4 \% → 78.7 \% for \textit{Qwen 2.5}/\textit{DiverseVul} and macro-F\textsubscript{1} from 0.54 → 0.65 for \textit{LLaMA 8B}/\textit{CleanVul}. GRPO also outperforms Supervised Fine-tuning in all settings and lifts macro-F\textsubscript{1} by 15–25 pp (e.g., 0.38 → 0.67 for \textit{Qwen 2.5}/\textit{DiverseVul}), increases accuracy to 0.58–0.67, and boosts the recall of \textit{Not Vulnerable} classfrom 0.07 → 0.61 (0.09 → 0.42 for \textit{LLaMA 8B}/\textit{CleanVul}); these gains persist for both in- and out-of-distribution data.}

To summarize, the contributions of this paper are:
\begin{itemize}
\item We extend previous analysis~\cite{FontanaSimoni2025} on understanding the performance of pretrained small instruction-tuned LLMs for vulnerability detection. We show that, even when explicitly prompted to reason, these models fail to effectively leverage their reasoning capabilities, leading to unsatisfactory outcomes.

\item We enable and apply self-reasoning during training through Group Relative Policy Optimization for the vulnerability detection task. This is achieved by introducing a modular, task-specific reward function tailored for vulnerability detection.

\item We conduct an extensive suite of experiments to demonstrate the benefits of the proposed GRPO formulation, showing consistent improvements, with respect on-the-self LLMs and SFT, in both prediction accuracy and reasoning quality. 
\end{itemize}

The remainder of this paper is structured as follows: Section~\ref{sec:background} provides an overview of software vulnerability detection and introduces the GRPO algorithm. Section~\ref{sec:zero-shot} examines the capabilities of instruction-tuned LLMs in zero-shot settings. Section~\ref{sec:GRPO} investigates the improvements achieved by training LLMs with GRPO. Section~\ref{sec:GRPOvsSFT} compares GRPO with supervised finetuning. Section~\ref{sec:ablation} presents ablation studies on the use of GRPO. Section~\ref{sec:related} reviews related work and adresses the proposed analysis within the existing literature.
Finally, Section~\ref{sec:conclusion} outlines the conclusions and future directions.

\section{Background and Preliminaries}

\label{sec:background}
This section provides an overview of the core background concepts of the work, including the challenges of automatic vulnerability detection and the utilization and training strategies of recent LLMs. It also introduces the experimental settings, which are consistently applied throughout the evaluations and analyses in the following sections.
%and the different finetuning strategies for instruction-based LLMs that we can adopt for vulnerability detection task.

\subsubsection{Software Vulnerability Detection}
Automated vulnerability detection aims to identify flaws like buffer overflows or logic errors that compromise system security. As software grows in complexity, manual review becomes unfeasible~\cite{hard-to-mantain-code}, and traditional tools struggle with evolving language features~\cite{too-many-languages}.
In this context, recent LLMs, pretrained on large codebases, have shown strong zero and few shot capabilities in spotting vulnerabilities~\cite{Analysing_LLM_Capabilities}, outperforming older techniques (as discussed in Section \ref{sec:related}). However, most LLM-based detectors still operate at the line level and lack mechanisms to track control or data flow across code sequences~\cite{Analysing_LLM_Capabilities}.
% Automated software vulnerability detection seeks to uncover flaws, such as buffer overflows, injection points, or logic errors, that attackers could exploit to undermine a system’s integrity, confidentiality, or availability. As codebases grow in size and complexity, manual code review becomes impractical~\cite{hard-to-mantain-code}, and even mature tools struggle to keep pace with new language features and idioms~\cite{too-many-languages}.

% Traditional static analysis methods inspect code without executing it, using heuristics or formal rules to flag suspicious constructs. They excel at catching simple patterns but often generate high rates of false positives and miss context‑sensitive flaws~\cite{high-fp-static}. Dynamic analysis (fuzzing) executes programs on crafted inputs to observe failures at runtime; while powerful for memory errors, it can fail to reach deep code paths or reason about semantic bugs~\cite{static_analysis_bad, static_analysis_bad_2}.

% More recently, large language models (LLMs) pretrained on massive code repositories have shown strong zero‑ and few‑shot ability to recognize vulnerable patterns with minimal feature engineering~\cite{outside-confort}. Despite their syntactic prowess, LLM-based detectors generally make isolated, line by line predictions and do not provide a structured way to follow a program’s control or data flow in sequence \cite{Analysing_LLM_Capabilities}.

\subsection{Finetuning strategies for instruction-based LLMs.}
The analysis proposed in the paper address three different strategies for utilizing LLMs in vulnerability detection: (i) no finetuning (zero-shot prompt engineering); (ii) supervised finetuning; and (iii) finetuning via Group Relative Policy Optimization.

\noindent \textbf{Zero-shot Prompt Engineering} consists in evaluating a pre-trained instruction-tuned LLM directly through natural language prompts, without updating its parameters (no additional finetuning). In Section~\ref{sec:zero-shot}, we compare two variants: a \textit{reasoning prompt}, which asks the model to justify its answer step by step, and a \textit{direct prompt}, which elicits only a final yes/no verdict. Since there is no universally optimal prompt format, we design both prompts from scratch to reflect the two extremes of explicit reasoning and concise decision-making. This strategy serves as a baseline to assess the model’s out-of-the-box reasoning capabilities.

%\subsection{Supervised Fine-tuning (SFT)}

\noindent \textbf{Supervised Finetuning (SFT)} is a widely used technique to specialize pretrained language models for downstream tasks by training them on curated input-output pairs. In the context of software vulnerability detection, SFT involves exposing the model to annotated code snippets in which vulnerabilities are labeled, enabling it to learn mappings from code patterns to vulnerability types or likelihoods~\cite{Harnessing_LLMs, LLMAO}.
SFT relies on direct supervision in the form of ground-truth labels, which makes it effective for initial task adaptation. However, when used alone, it may be limited by the quality and coverage of the labeled dataset~\cite{sft-problem-noise-dataset}, potentially leading to overfitting on common examples while missing subtle or novel flaws. Furthermore, SFT typically lacks mechanisms to encourage exploration or nuanced reasoning over complex structures, such as source code~\cite{reasoning-bad-sft}.

\noindent \textbf{Group Relative Policy Optimization (GRPO)} is a reinforcement learning algorithm proposed by DeepSeek~\cite{GRPO, GRPO-r1}, designed to improve the stability of training and reduce overfitting compared to SFT, using policy updates similar to Proximal Policy Optimization (PPO)~\cite{schulman2017proximal}. The core idea is to leverage multiple sampled outputs during training, generated in response to the same input prompt, to estimate and apply rewards in a relative manner. 
Formally, to train instruction-based LLMs, the LLMs is trained with a GRPO loss, denoted as $\mathcal{J}_{\text{GRPO}}(\theta)$:

\small
{
\begin{equation} 
\mathbb{E}_{q \sim \mathbb{P}(Q),\, \{o_i\} \sim \pi_{\theta_{\text{old}}}} \left[
\frac{1}{G} \sum_{i=1}^{G} \frac{1}{|o_i|} \sum_{t=1}^{|o_i|} 
\mathcal{C}_{i,t} - \beta\, D_{\text{KL}}\left( \pi_\theta \,\|\, \pi_{\text{ref}} \right)
\right]
\label{eq:grpo-full}
\end{equation}
}

\normalsize
\noindent
where, $\mathbb{P}(Q)$ is the distribution over input queries $q$, and $\pi_{\theta_{\text{old}}}$ is the policy used to sample $G$ model answers $\{o_i\}$. Each $o_i$ has length $|o_i|$ and $\mathcal{C}_{i,t}$ denotes the clipped policy advantage, computed as:
$
\mathcal{C}_{i,t} = \min\left( \frac{\pi_\theta(a_t|s_t)}{\pi_{\theta_{\text{old}}}(a_t|s_t)} \hat{A}_{i,t}, \text{clip} \right)
$,
where $\hat{A}_{i,t}$ is the estimated advantage at time step $t$ for the $i$-th sampled output $o_i$, and \textit{clip} represents a threshold as in PPO-style updates~\cite{schulman2017proximal}.  The ratio $\tfrac{\pi_\theta}{\pi_{\theta_{\text{old}}}}$ captures the relative likelihood of the current policy $\pi_\theta$ compared to the old policy. The final term applies a Kullback–Leibler divergence penalty ($D_{KL}$) between the current $\pi_\theta$ and a reference policy $\pi_{\rm ref}$, scaled by $\beta$.

Training models with this approach can be time consuming and resource intensive. Therefore, a simplified version of the original loss is used in this work. Specifically, following common practice, we adopt a simplified setting in which a single iteration is performed per training step, that is, each training sample is seen only once during a training epoch.
Under this setup, at the first iteration, the policy ratio is always $1$ since $\pi_\theta = \pi_{\theta_{\text{old}}}$, making the clipping operation unnecessary. To mitigate the \textit{question-level difficulty bias}~\cite{liu2025understanding}, we define the advantage as $\hat{A}_{i,t} = R_i - \bar{R}$ instead of normalizing by the standard deviation, where $R_i$ is the reward for the $i$-th model response and $\bar{R}$ is the mean reward within the group $G$.
Finally, we set $\beta = 10^{-6}$ for the main experiments, while ablation studies on its effect are presented in Section~\ref{sec:ablation}. The resulting GRPO objective simplifies to:

\begin{equation}
\mathcal{J}_{\text{GRPO}}(\theta) = \frac{1}{G} \sum_{i=1}^{G} (R_i - \bar{R}) - 10^{-6} \cdot \mathrm{D}_{\mathrm{KL}} \left[ \pi_{\theta} \| \pi_{\mathrm{ref}} \right]
\label{eq:grpo-simple}
\end{equation}

This formulation, originally developed and evaluated in the context of complex reasoning tasks such as mathematical problem solving~\cite{GRPO}, has demonstrated strong performance in optimizing LLMs and addressing challenges encountered with earlier policy-based techniques and SFT~\cite{grpo-reward-free}.
However, applying GRPO to a new task, such as vulnerability detection, requires the definition of suitable reward and advantage functions, which has not yet been explored previously in this context.

% However, using both outcome-level and step-level supervision~\cite{grpo-reward-free}. 
% QUI GIUSTIFICARE COSA MANCA E AFFRONTIAMO IN QUESTO LAVORO

%In this paper, we explore the potential of GRPO in a new application domain: vulnerability detection, where an LLM must navigate and assess code structures effectively to identify security flaws.

\subsection{Setup and Settings}
\label{sec:data-met}
\label{subsec:vul_data}

% METTERE CON BACKGROUN AND SETUP OF THE ANALYSIS INSIEME

\paragraph{Models} Throughout the analysis and experiments proposed in this paper, we evaluate three representative small-side LLMs, such as \texttt{LLaMA 8B}~\cite{llama3}, \texttt{LLaMA 3B}~\cite{llama3}, and \texttt{Qwen 2.5 3B}~\cite{yang2025qwen3}. These models were chosen due to their open-source availability, good performance and widespread adoption, making them ideal for research contexts where computational efficiency is critical. Note that, differently from other works~\cite{ProRLearn, LLMAO}, we address more generic  models that are not specifically pretrained or finetuned on code, because we want to exploit the abilities of these models to reason about codes so they should be better at reasoning than coding.

\paragraph{Datasets} To evaluate these models, we use three vulnerability datasets, \textit{DiverseVul}, \textit{BigVul}, and \textit{CleanVul}, each selected for its specific strengths in training and evaluation. In particular, \textit{DiverseVul} contains 18,945 vulnerable functions across 150 CWEs, along with 330,492 non-vulnerable functions. Its high level of manual curation ensures high label precision, making it well-suited for model finetuning~\cite{diversevul}; \textit{BigVul} is a large-scale dataset of C/C++ vulnerabilities, compiled by crawling the public CVE database and associated GitHub repositories~\cite{bigvul}; \textit{CleanVul} comprises 8,203 code functions from six languages (Java, Python, JavaScript, C\#, C, and C++). It was created using an LLM-based heuristic pipeline (achieving 90.6\% overall correctness) and is primarily designed to assess model generalization across languages. Due to the nature of its construction, it is not possible to associate each snippet with a specific CWE~\cite{cleanvul}.
In both Section~\ref{sec:GRPO} and \ref{sec:GRPOvsSFT}, we use 80\% of DiverseVul dataset to train our models and we let the other 20\% as test dataset.
To maintain consistency across experiments, all datasets were balanced and truncated such that the resulting prompts did not exceed the 4,000 token maximum input length.

% In Section \ref{sec:zero-shot}, we ask the model to say if is vulnerable or not letting the model to say this directly or after a reason about the code. In Section\ref{sec:GRPO}, we train the models using to reason about the provided code, using reinforcement learning (GRPO) and in Section\ref

\paragraph{Metrics} The goal here is to understand how these models perform in detecting vulnerabilities, that means, given a code in input to the large language model, it should answer \textit{Not Vulnerable} or \textit{Vulnerable} during the inference phase.
Hence, given this binary classification setting, where (\textit{Vulnerable} is the positive class and \textit{Not Vulnerable} is the negative class), we define the standard and well-known confusion matrix components for each class as follows: true positives (TP), false positives (FP), true negatives (TN), and false negatives (FN). Based on these, we can compute the class-wise performance metrics as: 
\small
\begin{equation}
\texttt{Accuracy} = \frac{TP + TN}{TP + TN + FP + FN}
\end{equation}
\begin{equation}
\texttt{Precision} = \frac{TP}{TP + FP}, \qquad
\texttt{Recall} = \frac{TP}{TP + FN}
\end{equation}

\begin{equation}
\texttt{F1-Score} = \frac{2 \cdot \texttt{Precision} \cdot \texttt{Recall}}{\texttt{Precision} + \texttt{Recall}}
\end{equation}
\normalsize

To summarize overall model performance, we report in the subsequent analysis the following scores:
\begin{itemize}
    \item \textbf{Macro F1}: the unweighted mean of the F1 Scores across all classes, treating each class equally.
    \item \textbf{Weighted F1}: the mean of the F1 Scores weighted by the support (number of true instances) of each class.
\end{itemize}

\section{Reasoning Capabilities of Small LLMs}
\label{sec:zero-shot}

In this section, we investigate our first research question \RQ{RQ1}, which explores the zero-shot capabilities of small off-the-shelf language models for vulnerability detection. In particular, we evaluate the original (non finetuned) LLMs in their instruction-tuned baseline versions using two different zero-shot \textit{system prompts}, to explore the performance both in terms of detection accuracy and reasoning. 
In the first system prompt, we directly ask the model to say whether the provided code is vulnerable or not (first block in Figure~\ref{fig:prompts_comparison}) and in the second system prompt, we ask to reason about the provided code before making the final verdict (second block in Figure~\ref{fig:prompts_comparison}). In both cases, the \textit{system prompt} is followed by a \textit{user prompt} (that is the same between reasoning and no reasoning prompt setting) with the code snippet to be analyzed (bottom block in Figure~\ref{fig:prompts_comparison}).
 
\begin{figure}[ht]
  \centering
    \centering
    % --- prima figura ---
    \begin{tcolorbox}[colback=black!5!white,colframe=black!75!black,
      title={System Prompt - \textit{No Reasoning} Test}]
      \begin{minipage}{\columnwidth}
    \footnotesize
You are a senior software security auditor.

When I give you a code snippet, silently analyse it for any realistic, exploitable security weakness (e.g. injections, memory-safety errors, race conditions, logic flaws, privilege escalation, unsafe crypto, misuse of APIs, etc.).

\textit{Decision:}
\begin{itemize}
    \item If you identify at least one credible exploit path, reply with the single word  \textbf{ vulnerable}
    \item If you find none, reply with the single word   \textbf{not vulnerable}
\end{itemize}
\textit{Output rules (MUST follow):}
\begin{enumerate}
    \item  Reply with \textbf{exactly} one of the two words above-nothing else.
    \item  No additional text, quotes, punctuation, markdown, or whitespace before/after the word.
    \item Perform all reasoning internally; do \textbf{not} expose it in the output.
\end{enumerate} 
    \end{minipage}
    \end{tcolorbox}
    % --- seconda figura ---
    \begin{tcolorbox}[colback=black!5!white,colframe=black!75!black,
      title={System Prompt - \textit{Reasoning} Test}]
     \begin{minipage}{\columnwidth}
    \footnotesize
You are a software security expert. I will provide you with code snippets. Your task is to analyze the code and determine whether it contains any vulnerabilities.
\texttt{<reasoning>}
\begin{enumerate}
    \item  Summarize the code snippet.
    \item Check thoroughly for vulnerabilities.
    \item Clearly state whether the code is secure or insecure.
\end{enumerate}
\texttt{</reasoning>}
\texttt{<answer>}
\textbf{Yes, the code is vulnerable.}
\textit{OR}
\textbf{No, the code is not vulnerable.}
\texttt{</answer>}
    \end{minipage}
    \end{tcolorbox}
    %\caption{\small{\textit{Reasoning Test} - System Prompt}}
  \begin{tcolorbox}[colback=black!5!white,colframe=black!75!black,
title={User Prompt - \textit{Reasoning} and \textit{No Reasoning} Test}]
    \begin{minipage}{\columnwidth}
    \footnotesize
Analyze the following code snippet to determine whether it is vulnerable.
\texttt{{CODE SNIPPET}}
    \end{minipage}
\end{tcolorbox}%
  \caption{System prompts for the no-reasoning and reasoning settings (top and middle blocks, respectively), and the user prompt that is appended to both.}
  \label{fig:prompts_comparison}
\end{figure}
%------------- PROMPT ------------------------

%---------------------------------------

The rationale behind these experiments is that, by comparing results across both settings and multiple datasets, we can better understand how small language models behave when making security-critical decisions, and whether reasoning or \textit{thinking aloud} helps in detection metrics. %reducecommon error patterns.
Note that, in the reasoning prompt, each model must provide its own final verdict enclosed within \texttt{<answer>} and \texttt{</answer>}, which is then compared against the ground-truth annotations provided by the datasets.

\begin{table*}[htbp]
\centering
\tiny
\caption{Comparison across models (Qwen 2.5, LLaMA 3B, LLaMA 8B) on datasets: CleanVul, DiverseVul, and BigVul — Without Reasoning (NR) vs With Reasoning (R)}
\setlength{\tabcolsep}{1pt}
\renewcommand{\arraystretch}{0.9}
\rowcolors{3}{gray!5}{white}
\resizebox{\textwidth}{!}{%
\begin{tabular}{l|cc|cc|cc| cc|cc|cc| cc|cc|cc}
    \toprule
    & \multicolumn{6}{c|}{\textbf{Qwen 2.5}}
    & \multicolumn{6}{c|}{\textbf{LLaMA 3B}}
    & \multicolumn{6}{c}{\textbf{LLaMA 8B}} \\
    & \multicolumn{2}{c}{\textcolor{blue}{\textbf{CleanVul}}}
    & \multicolumn{2}{c}{\textcolor{green!50!black}{\textbf{DiverseVul}}}
    & \multicolumn{2}{c|}{\textcolor{orange!80!black}{\textbf{BigVul}}}
    & \multicolumn{2}{c}{\textcolor{blue}{\textbf{CleanVul}}}
    & \multicolumn{2}{c}{\textcolor{green!50!black}{\textbf{DiverseVul}}}
    & \multicolumn{2}{c|}{\textcolor{orange!80!black}{\textbf{BigVul}}}
    & \multicolumn{2}{c}{\textcolor{blue}{\textbf{CleanVul}}}
    & \multicolumn{2}{c}{\textcolor{green!50!black}{\textbf{DiverseVul}}}
    & \multicolumn{2}{c}{\textcolor{orange!80!black}{\textbf{BigVul}}} \\
    & \cellcolor{green!20}\textbf{NR$^\star$} & \textbf{R}
    & \cellcolor{green!20}\textbf{NR$^\star$} & \textbf{R}
    & \cellcolor{green!20}\textbf{NR$^\star$} & \textbf{R}
    & \textbf{NR$^\star$} & \cellcolor{green!20}\textbf{R}
    & \textbf{NR} & \cellcolor{green!20}\textbf{R$^\star$}
    & \textbf{NR} & \cellcolor{green!20}\textbf{R$^\star$}
    & \textbf{NR} & \cellcolor{green!20}\textbf{R$^\star$}
    & \textbf{NR} & \cellcolor{green!20}\textbf{R$^\star$}
    & \textbf{NR} & \cellcolor{green!20}\textbf{R$^\star$} \\
    \cmidrule(r){2-7}\cmidrule(r){8-13}\cmidrule(r){14-19}
    \rowcolor{gray!15}\multicolumn{19}{l}{\textbf{Not Vulnerable}} \\
    Precision         & \textbf{0.56} & 0.52 & \textbf{0.53} & 0.49 & \textbf{0.56} & 0.54
                     & \textbf{1.00} & 0.53 & \textbf{0.67} & 0.46 & 0.00 & \textbf{0.52}
                     & 1.00 & \textbf{0.50} & 0.00 & \textbf{0.44} & \textbf{1.00} & 0.51 \\
    Recall            & 0.65 & \textbf{0.80} & 0.55 & \textbf{0.75} & \textbf{0.91} & 0.87
                     & 0.00 & \textbf{0.47} & 0.00 & \textbf{0.39} & 0.00 & \textbf{0.53}
                     & 0.00 & \textbf{0.92} & 0.00 & \textbf{0.26} & 0.00 & \textbf{0.96} \\
    F\textsubscript{1} & 0.60 & \textbf{0.63} & 0.54 & \textbf{0.59} & \textbf{0.69} & 0.66
                     & 0.00 & \textbf{0.50} & 0.00 & \textbf{0.42} & 0.00 & \textbf{0.52}
                     & 0.00 & \textbf{0.65} & 0.00 & \textbf{0.33} & 0.00 & \textbf{0.67} \\
    \hline
    Support & \multicolumn{2}{c|}{6330} & \multicolumn{2}{c|}{2952} & \multicolumn{2}{c|}{1134} & \multicolumn{2}{c|}{6330} & \multicolumn{2}{c|}{2952} & \multicolumn{2}{c|}{1134} & \multicolumn{2}{c|}{6330} & \multicolumn{2}{c|}{2952} & \multicolumn{2}{c|}{1134} \\
    \midrule
    \rowcolor{gray!15}\multicolumn{19}{l}{\textbf{Vulnerable}} \\
    Precision         & \textbf{0.59} & 0.56 & \textbf{0.52} & 0.43 & \textbf{0.75} & 0.62
                     & 0.50 & \textbf{0.52} & \textbf{0.49} & 0.45  & 0.49 & \textbf{0.50}
                     & 0.50 & \textbf{0.53} & \textbf{0.49} & 0.45 & 0.49 & \textbf{0.55} \\
    Recall            & \textbf{0.49} & 0.26 & \textbf{0.50} & 0.19 & \textbf{0.27} & 0.22
                     & \textbf{1.00} & 0.58 & \textbf{1.00} &  0.52 & \textbf{1.00} & 0.50
                     & \textbf{1.00} & 0.09 & \textbf{1.00} & 0.65 & \textbf{1.00} & 0.06 \\
    F\textsubscript{1} & \textbf{0.53} & 0.35 & \textbf{0.51} & 0.26 & \textbf{0.40} & 0.33
                     & \textbf{0.67} & 0.55 & \textbf{0.66} & 0.48 & \textbf{0.66} & 0.50
                     & \textbf{0.67} & 0.15 & \textbf{0.66} & 0.54 & \textbf{0.66} & 0.10 \\
    \hline
    Support & \multicolumn{2}{c|}{6330} & \multicolumn{2}{c|}{2952} & \multicolumn{2}{c|}{1134} & \multicolumn{2}{c|}{6330} & \multicolumn{2}{c|}{2952} & \multicolumn{2}{c|}{1134} & \multicolumn{2}{c|}{6330} & \multicolumn{2}{c|}{2952} & \multicolumn{2}{c|}{1134} \\
    \midrule
    \rowcolor{gray!15}\multicolumn{19}{l}{\textbf{Overall}} \\
    Accuracy           & \textbf{0.57} & 0.53 & \textbf{0.52} & 0.48 & \textbf{0.59} & 0.55
                      & 0.50 & \textbf{0.53} & \textbf{0.49} &  0.45 & 0.49 & \textbf{0.51}
                      & 0.50 & \textbf{0.50} & \textbf{0.49} & 0.45 & 0.49 & \textbf{0.51} \\
    Macro Precision    & \textbf{0.57} & 0.54 & \textbf{0.52} & 0.46 & \textbf{0.65} & 0.58
                      & \textbf{0.75} & 0.53 & \textbf{0.58} & 0.45 & 0.25 & \textbf{0.51}
                      & \textbf{0.75} & 0.52 & 0.25 & \textbf{0.45} & \textbf{0.75} & 0.53 \\
    Macro Recall       & \textbf{0.57} & 0.53 & \textbf{0.52} & 0.47 & \textbf{0.59} & 0.55
                      & 0.50 & \textbf{0.53} & \textbf{0.50} & 0.46 & 0.50 & \textbf{0.51}
                      & 0.50 & \textbf{0.50} & \textbf{0.50} & 0.45 & 0.50 & \textbf{0.51} \\
    Macro F\textsubscript{1} & \textbf{0.57} & 0.49 & \textbf{0.52} & 0.43 & \textbf{0.54} & 0.50
                            & 0.33 & \textbf{0.53} & 0.33 & \textbf{0.45}  & 0.33 & \textbf{0.51}
                            & 0.33 & \textbf{0.40} & 0.33 & \textbf{0.43} & 0.33 & \textbf{0.38} \\
    Weighted Precision & \textbf{0.57} & 0.54 & \textbf{0.52} & 0.46 & \textbf{0.65} & 0.58
                      & \textbf{0.75} & 0.53 & \textbf{0.58} & 0.45 & 0.24 & \textbf{0.51}
                      & \textbf{0.75} & 0.52 & 0.24 & \textbf{0.45} & \textbf{0.75} & 0.53 \\
    Weighted Recall    & \textbf{0.57} & 0.53 & \textbf{0.52} & 0.48 & \textbf{0.59} & 0.55
                      & 0.50 & \textbf{0.53} & \textbf{0.49} & 0.45 & 0.49 & \textbf{0.51}
                      & 0.50 & \textbf{0.50} & \textbf{0.49} & 0.45 & 0.49 & \textbf{0.51} \\
    Weighted F\textsubscript{1} & \textbf{0.57} & 0.49 & \textbf{0.52} & 0.43 & \textbf{0.55} & 0.50
                               & 0.33 & \textbf{0.53} & 0.33 & \textbf{0.45} & 0.33 & \textbf{0.51}
                               & 0.33 & \textbf{0.40} & 0.33 & \textbf{0.43} & 0.33 & \textbf{0.39} \\ 
                               \hline
    Support & \multicolumn{2}{c|}{12660} & \multicolumn{2}{c|}{5904} & \multicolumn{2}{c|}{2268} & \multicolumn{2}{c|}{12660} & \multicolumn{2}{c|}{5904} & \multicolumn{2}{c|}{2268} & \multicolumn{2}{c|}{12660} & \multicolumn{2}{c|}{5904} & \multicolumn{2}{c|}{2268} \\
    \bottomrule
\end{tabular}}
\label{tab:comparison-R-NR}
\end{table*}

Table~\ref{tab:comparison-R-NR} shows clear differences in how models respond to reasoning.  
Qwen 2.5 works best without reasoning: it achieves the highest overall scores across all datasets, with a macro F\textsubscript{1} of 0.57 on CleanVul, 0.52 on DiverseVul, and 0.54 on BigVul. When reasoning is added, its performance drops, especially for the \texttt{Vulnerable} class, where recall falls by half or more. This suggests that Qwen already has effective internal strategies for detecting vulnerabilities, and forcing it to explain its answers can introduce confusion and reduce accuracy.

The opposite happens with LLaMA models. Without reasoning, both LLaMA 3B and LLaMA 8B predict almost everything as \texttt{Vulnerable}. This leads to perfect recall on that class, but zero recall on \texttt{Not Vulnerable}, which hurts overall performance. Adding reasoning helps these models balance their predictions. For example, LLaMA 3B’s F\textsubscript{1} score on the \texttt{Not Vulnerable} class goes from 0.00 to 0.50 on CleanVul and reaches 0.52 on BigVul. Its macro F\textsubscript{1} improves from 0.33 to over 0.45 in all datasets. LLaMA 8B shows the same trend, with the biggest gain on CleanVul where the F\textsubscript{1} jumps from 0.00 to 0.65 on \texttt{Not Vulnerable} class. However, this improvement comes at the cost of lower recall on the \texttt{Vulnerable} class. For instance, on BigVul, LLaMA 8B’s recall drops from 1.00 to just 0.06 when reasoning is added.

Overall, reasoning helps LLaMA models make better distinctions between vulnerable and not vulnerable code, especially on harder datasets like BigVul. In contrast, Qwen performs better when it makes predictions directly, without explaining them. This suggest that whether reasoning helps or not depends on the model and the goal: if the task needs fewer false positives, reasoning is useful for LLaMA; but if high vulnerability coverage is more important, Qwen without reasoning may be the better choice. To summarize, based on the proposed analysis, we conclude the following:

\begin{tcolorbox}[colback=black!1!white,colframe=black!1!black]
    \begin{minipage}{\columnwidth}
\RQ{RQ1} \textit{Can a small instruction tuned LLM reason effectively about software vulnerabilities without additional finetuning?}

\RQ{A1} \textit{Experimental results indicate that models struggle to fully leverage reasoning when analyzing software vulnerabilities. Even in the case of LLaMA, where \textit{thinking aloud} leads to improvements, its predictions remain unbalanced and prone to false negatives (e.g LLaMA 8b).}

\end{minipage}
\end{tcolorbox}%

\section{Improving Model Reasoning Abilities}
\label{sec:GRPO}

Previous results remark that, although allowing LLMs to reason about the code may improve performance, the models remain unreliable. To further enhance the ability to detect vulnerabilities more consistently, it is necessary to train the model to improve the quality of its own reasoning. To this end, in this section we explore the possibility of adapting and applying reasoning-aware policies during training for more effective finetuning of LLMs \RQ{RQ2}.

In this context, as known in the literature, there are no available datasets that support direct finetuning of models to also reason about vulnerabilities based on ground truth step by step exaplanations, as existing datasets only provide binary ground truth labels. This limitation can be addressed through RL-based approaches, such as GRPO, which allow models to generate their own reasoning while policy parameters are optimized based on predefined reward signals.
To enable this, before adopting GRPO in our training pipeline, it is necessary to define appropriate reward functions tailored to vulnerability detection.

%the problem of finetuning of model by teaching the reasoning 

% The results of Table~\ref{tab:comparison-R-NR} shows that, even if letting the LLM reason about the code boost the ability of these models (at least for LLaMA architectures), they are still quite unrealiable. So, in order to improve the model ability to detect vulnerabilities, we need to train the model to improve the quality of its own reasoning. 

% Since, no dataset available that we can use to fine-tune the models to reason on the vulnerability detection task in a supervised way because the available datasets do not provide explanations or reasonings on the code they collected, but they provide only the final ground truth label (\textit{vulnerable} or \textit{not vulnerable}).
% This issue can be overcome with the help of reinforcement learning, through which we can let models generate reasoning and give their final verdict and at the end we can reshape policy parameters using predefined rewards.

\subsection{Reward Design}
To improve the model's ability to detect vulnerabilities by using its own reasoning, we should not only give high rewards to responses that make the correct verdict, but we also need to train the model to format the answers well. 

\begin{figure}[ht]
  \includegraphics[width=0.5\textwidth]{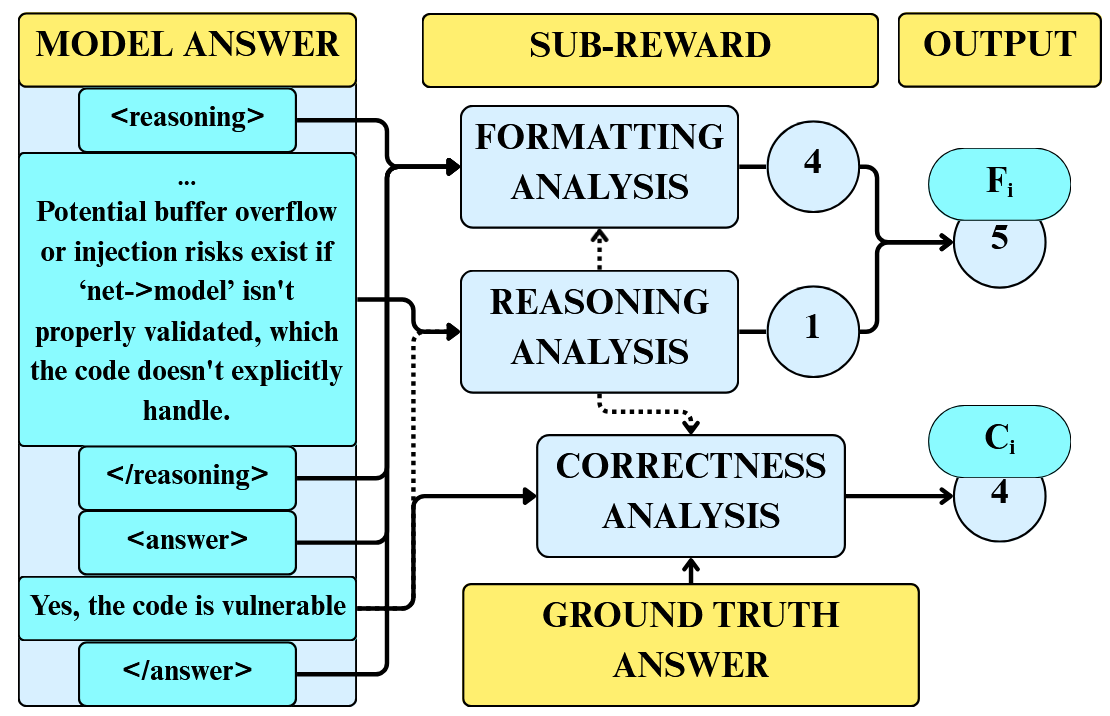}
  \caption{\textit{Sub-Rewards} signals generated by \textit{Formatting, Reasoning and Correctness Analyses}. The dotted lines indicate that incoherent reasoning by the model also nullifies the rewards from correctness and formatting.}
  \label{fig:ini-reward}
\end{figure}

We design a modular reward function that computes the final reward signal by performing three distinct analyses (see Figure~\ref{fig:ini-reward}): \textit{Formatting Analysis}, \textit{Correctness Analysis}, and \textit{Reasoning Analysis}. Each of these analyses produces a corresponding \textit{sub-reward}. For each model answer $i$ belonging to the set of model answers $G$, the \textit{Formatting Analysis} and \textit{Reasoning Analysis} are aggregated to produce the sub-reward $F_i$, while the \textit{Correctness Analysis} produces a separate sub-reward $C_i$. The values $F_i$ and $C_i$ are then combined through a specific strategy that will be detailed later. To train the model we used the reasoning system prompt (second box) and the user prompt (third box) both shown in Figure~\ref{fig:prompts_comparison}. Below, we describe each of the analysis in detail:

\medskip
\noindent \textbf{Formatting analysis.}
In this step, we verify that the model adheres to the structure requested by the reasoning system prompt. Specifically, we check for the presence of the \texttt{<reasoning>} and \texttt{<answer>} tags and confirm that the reasoning follows the three-step pattern highlighted in second block of Figure~\ref{fig:prompts_comparison} (\textit{1.~Summarize the code snippet, 2.~Check for vulnerabilities, 3.~State if the code is secure or not}). If all these conditions are met, the model receives a reward of 4; if the tags are missing, the model receives a reward of 0.

\medskip
\noindent \textbf{Reasoning analysis.} 
We evaluate the quality of the explanation by assigning a bonus based on its length and lexical diversity. Let \( n \) be the number of tokens in the reasoning text. We compute a log-scaled length bonus as:

% \begin{equation}
% \text{LengthBonus}(n) = 5 \cdot \frac{\frac{1.4 \cdot \log(n+1)}{\log(2001)}}{1.4},
% \label{eq:length-bonus}
% \end{equation}

\begin{equation}
\text{LengthBonus}(n) = 5 \cdot \frac{ \log(n+1)}{\log(2001)},
\label{eq:length-bonus}
\end{equation}

which grows logarithmically with \( n \), and saturates as the token count approaches 2000. This discourages overly long  explanations. We also assess the final reasoning step (\textit{Step 3} in reasoning system prompt) by two metrics: the word-level edit distance to the model’s final answer contained in \texttt{<answer>} tags (using difflib\footnote{\url{https://docs.python.org/3/library/difflib.html}}) and a \textit{Coherence Score} based on cosine similarity of MiniLM-L6-v2 embeddings\footnote{\url{sentence-transformers/all-MiniLM-L6-v2}}. If an explanation merely restates the answer without adding insight, we apply a penalty (-2); if it instead adds useful information, we assign a bonus (+1). If its Coherence Score falls below 0.4, we mark the explanation as \textit{incoherent}. If \textit{incoherent} the total reward is 0.

    % We then check the explanation inside the \texttt{<reasoning>} tag and assign a reward based on two criteria:
    % \begin{itemize}
    
    %     \item \textbf{Length and wording variety.}  
    %     We reward explanations that are longer but not repetitive. We use a log-based bonus that grows with the number of words, but we reduce the reward if the explanation repeats the same words too much.

    %     \item \textbf{Consistency with the answer.}
    %     We assess the final reasoning step by two metrics: the word-level edit distance to the model’s final answer (using difflib\footnote{\url{https://docs.python.org/3/library/difflib.html}}) and a Coherence Score based on cosine similarity of MiniLM-L6-v2 embeddings\footnote{\url{sentence-transformers/all-MiniLM-L6-v2}}. If an explanation merely restates the answer without adding insight, we apply a small penalty; if its Coherence Score is sufficiently high, we award a modest bonus, but if the score falls below 0.4, we mark the explanation as \textit{incoherent}. If \textit{incoherent} the total reward of the model is 0.
    % \end{itemize}

\medskip
\noindent \textbf{Correctness analysis.}
For each model answer, we extract the content inside the \texttt{<answer>} tag and compare it to the correct answer, ignoring case. The answer is expected in the form ``\textit{Yes, the code is vulnerable}'' or ``\textit{No, the code is not vulnerable}''.  
If the answer exactly matches the correct answer, we assign a reward of 4. Otherwise, the reward is 0.

\medskip
\noindent \textbf{Dynamic rewards sum to avoid reward hacking.}
After generating the sub-rewards $C_i$ (correctness) and $F_i$ (formatting and reasoning), a proper strategy is required to combine them into a single reward value before being processed by the GRPO loss, which expects a single reward term. In this regard, we observed that following a naive approach that simply sums the two sub-rewards, i.e., \( r_i = 0.5 \cdot {F}_i + 0.5 \cdot {C}_i \) (denoted in the following plots as \textit{static reward}), leads to unsatisfactory performance during GRPO training.  
In fact, as shown in Table~\ref{tab:grpo-static}, both the LLaMA 8B and Qwen 2.5 models tend to label every code snippet as \textit{vulnerable}, even though they learn to format their responses correctly. The rationale behind this behavior is that vulnerability detection is a challenging task for small language models, which encourages them to exploit the reward signal by consistently defaulting to a single verdict. This reveals a clear instance of \textit{reward hacking}~\cite{wang2025beyond, reward_hacking}, where the model achieves a neutral advantage by repeating the same answer regardless of correctness\footnote{In GRPO, this phenomenon occurs when all model answers are either entirely correct or entirely incorrect. In both cases, the lack of variance in the advantage signal prevents meaningful learning and instead rewards non-informative or trivial behaviors.}.
% So at this point, one can ask \textit{is the model able to improve its own reasoning and detection capabilities by simply using the signal from the sum of $C_i$ and $F_i$?} As can be seen in Table~\ref{tab:grpo-static}, the results are still unsatisfactory: both the Llama 8B and Qwen 2.5 models will always mark each code as \textit{vulnerable}, even if they learned to format the answers correctly. Vulnerability detection is a difficult task for small language models, which leads them to exploit the reward signal by defaulting to responses with the same final verdict. In fact, this scenario reveals a clear instance of \textit{reward hacking}~\cite{wang2025beyond, reward_hacking}: if the model always outputs the same answer, regardless of its correctness, it can still obtain an advantage of zero for each model answer. In GRPO, the same outcome occurs even when all model answers are entirely correct or entirely incorrect. In both cases, the lack of variance in the advantage signal provides no meaningful learning signal, effectively rewarding non-informative or trivial behaviors.
% dynamic reward module - andiamo a sommari il segnale in maniera dipende da alpha (che guarda se il formatting è stabile oppure no). 

\begin{table}[htbp]
    \centering
    \small
    \setlength{\tabcolsep}{5pt}
    \renewcommand{\arraystretch}{1.1}
    \begin{tabular}{l|c|c}
        \toprule
        \textbf{Metric} & \textbf{LLaMA 8B} & \textbf{Qwen 2.5} \\
        \midrule
        \rowcolor{gray!15} \multicolumn{3}{l}{\textbf{Not Vulnerable}} \\
        Precision            & 0.00 & 0.00 \\
        Recall               & 0.00 & 0.00 \\
        F\textsubscript{1}-Score       & 0.00 & 0.00 \\
        \hline
    Support & \multicolumn{2}{c}{2952}\\
        \midrule
        \rowcolor{gray!15} \multicolumn{3}{l}{\textbf{Vulnerable}} \\
        Precision            & 0.50 & 0.50 \\
        Recall               & 1.00 & 1.00 \\
        F\textsubscript{1}-Score       & 0.67 & 0.67 \\
        \hline
    Support & \multicolumn{2}{c}{2952}\\
        \midrule
        \rowcolor{gray!15} \multicolumn{3}{l}{\textbf{Overall}} \\
        Accuracy             & 0.50 & 0.50 \\
        Macro Precision      & 0.25 & 0.25 \\
        Macro Recall         & 0.50 & 0.50 \\
        Macro F\textsubscript{1}-Score & 0.33 & 0.33 \\
        Weighted Precision   & 0.25 & 0.25 \\
        Weighted Recall      & 0.50 & 0.50 \\
        Weighted F\textsubscript{1}-Score & 0.33 & 0.33 \\
        \hline
        Support & \multicolumn{2}{c}{5904}\\
        \bottomrule
    \end{tabular}
     \caption{Results when using static reward for GRPO,  i.e., \( r_i = 0.5 \cdot {F}_i + 0.5 \cdot {C}_i \), on the DiverseVul dataset}        \label{tab:grpo-static}
\end{table}

To overcome this issue is necessary to build a reward signal that adapts over time: during the initial training phases, it is important that the model learns to follow the required format; once formatting is consistently correct and learned, the focus should gradually shift toward improving the quality of reasoning to enhance detection performance. 

\begin{figure}[t]
\centering
  \includegraphics[width=0.49\textwidth]{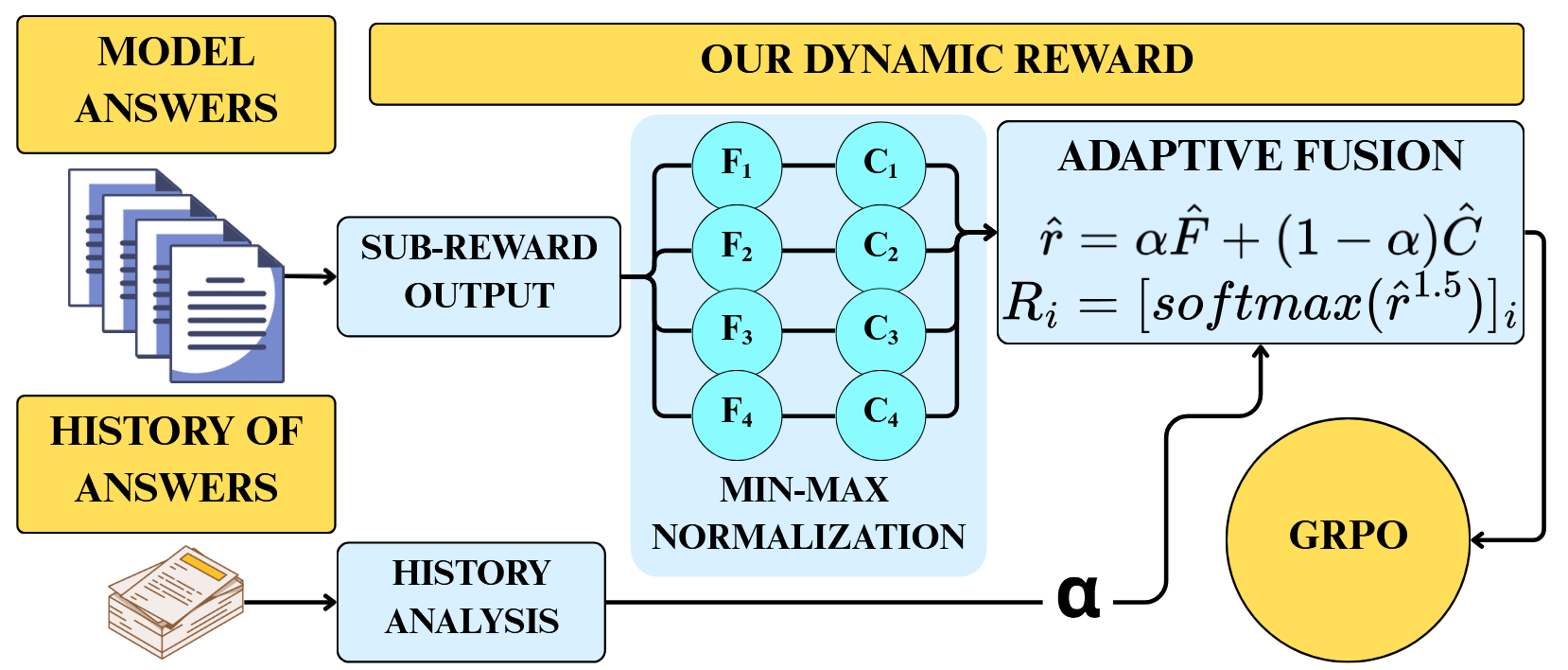}
  \caption{Final reward signal used to feedback the models}
  \label{fig:our-reward}
\end{figure}

To support this adaptive training process, we introduce a \textit{Dynamic Reward Module} (illustrated in Figure~\ref{fig:our-reward}), which adjusts the weighting of the sub-rewards $C_i$ and $F_i$ based on the model’s current ability to produce well formatted outputs.
In particular, each $C_i$ and $F_i$ scores are min–max normalized within each set of model answers $G$, and then combined as:
\begin{equation}
     r_i = \alpha \, \hat{F}_i + (1 - \alpha)\, \hat{C}_i,
     \label{eq:reward_ri}
\end{equation}
\noindent where $\hat{F}_i$ and $\hat{C}_i$ denote the normalized values of $F_i$ and $C_i$, respectively and $\alpha$ is \textit{dynamic} coefficient defined as:

% \small
% \begin{equation}
% \alpha \;=\;
% \begin{cases}
% 0.9, & \text{if } |H|<3 \;\text{or}\; \bar{F}_{H}\le 3,\\[6pt]
% \displaystyle\max\!\bigl(0.2,\,
%       \min\!\bigl(0.9,\;0.9 - 0.2\,\bar{F}_{H}\bigr)\bigr),
%       & \text{otherwise},
% \end{cases}
% \label{eq:alpha}
% \end{equation}
% \normalsize

\small
\begin{equation}
\alpha \;=\;
\begin{cases}
\tau_A, & \text{if } |H|<3 \;\text{or}\; \bar{F}_{H}\le 3,\\[6pt]
\displaystyle\max\!\bigl(\tau_B,\,
      \min\!\bigl(\tau_A,\tau_A- \,\tau_B \cdot \bar{F}_{H}\bigr)\bigr),
      & \text{otherwise},
\end{cases}
\label{eq:alpha}
\end{equation}
\normalsize

\noindent where \( H \) is a buffer that stores the mean \textit{formatting analysis sub-reward} (with a maximum value of 4) over the last three groups of answers, $\tau_A \geq \tau_B$, and \( \bar{F}_{H} \) is its running average.  In our setup, we set \( \tau_A = 0.9 \) and \( \tau_B = 0.2 \) based on a preliminary experimental exploration.  
Hence, training starts with \( \alpha = \tau_A \), placing strong emphasis on formatting. Once the model produces well-formatted outputs for three consecutive batches (i.e., \( \bar{F}_{H} > 3 \)), \( \alpha \) is gradually reduced, never falling below \( \tau_B \), so that correctness gains increasing weight in the reward signal.

%Next, we apply \textit{power scaling} to reinforce confident predictions: \( r_i \rightarrow r_i^{P_{\text{dyn}}} \) where the exponent $P_{\text{dyn}}$, kept fixed at 1.5, increases as the standard deviation of the correctness scores in the batch decreases.
Next, we apply \textit{power scaling} to reinforce confident predictions: \( r_i \rightarrow r_i^{1.5} \), using a fixed exponent to emphasize predictions with higher confidence. 
And apply the softmax function of all the answers $i$ accross the set $G$. 
Putting everything together, the final reward for model answer $i$ is computed as:
\small
\begin{equation}
R_i =
\begin{cases}
0, & \text{if } i \text{ is incoherent}, \\[4pt]
\bigl[\operatorname{Softmax}\big((\alpha \hat{F} + (1 - \alpha) \hat{C})^{1.5}\big)\bigr]_i, & \text{otherwise},
\end{cases}
\label{eq:reward}
\end{equation}
\normalsize
\noindent where $R_i$ is the final reward used by the GRPO algorithm, and $\hat{F}$ and $\hat{C}$ are the vectors of normalized formatting and correctness scores across all model answers in the produce set $G$, and 

\begin{figure}[ht]
  \centering
  \includegraphics[width=0.45\textwidth]{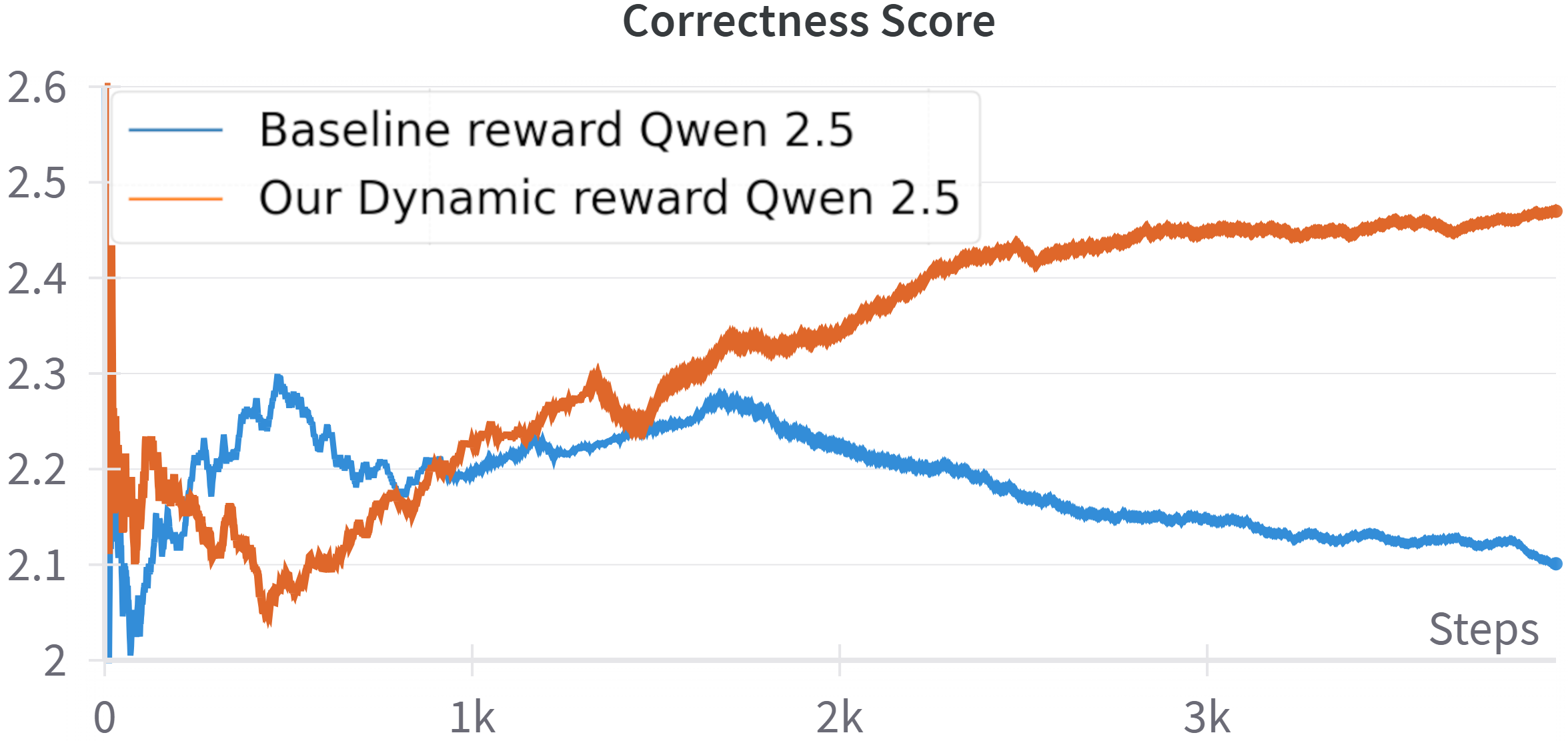}
  \caption{Running average of mean Correctness score ($\hat{C}$) during training, comparing our dynamic reward function to the traditional approach.}
  \label{fig:correctness-comparison}
\end{figure}

As illustrated in Figure~\ref{fig:correctness-comparison} (using Qwen 2.5 as a demonstrative example), this mechanism strengthens the reward signal for stable, high-confidence model answers and suppresses noisy or uncertain outputs. 
% The adaptive weighting strategy, which keeps \( \alpha \in [\tau_B, \tau_A] \), dynamically balances the emphasis between formatting and correctness based on recent model performance. Meanwhile, the use of \textit{power scaling} introduces a global, constant amplification of reward differences, further reinforcing learning stability.
In contrast, when using the static baseline sum of rewards, such as the mean correctness score (\( \text{mean}(\hat{C})_t \)). the value steadily increases throughout training.

\subsection{GRPO vs. Best Reasoning/No-Reasoning Baselines}
After introducing the proposed reward design, we first report comparisons between LLMs trained with our approach and the best-performing reasoning and no-reasoning zero-shot baselines that we can extract from Table~\ref{tab:comparison-R-NR}.

\begin{table}[ht]
  \centering
  \scriptsize
  \renewcommand{\arraystretch}{0.4}  % Reduce row height
  \rowcolors{3}{gray!10}{white}
  \resizebox{0.9\linewidth}{!}{%
    \begin{tabular}{l|l}
      \toprule
      \rowcolor{gray!25} \multicolumn{2}{l}{\textbf{GRPO Parameters}} \\
      $\beta$ for $D_{kl}$ & 1e-6 \\
      N. Model answers & 12 \\
      \midrule
      \rowcolor{gray!25} \multicolumn{2}{l}{\textbf{Training Setup}} \\
      N. train epochs & 1 \\
      Batch size & 4 \\
      Gradient accumulation steps & 4 \\
      Save steps & 500 \\
      \midrule
      \rowcolor{gray!25} \multicolumn{2}{l}{\textbf{Optimization Parameters}} \\
      Learning rate & 5e-6 \\
      Weight decay & 0.1 \\
      Adam beta1 & 0.9 \\
      Adam beta2 & 0.99 \\
      Optim & paged adamw 8bit \\
      LR scheduler type & cosine \\
      Warmup ratio & 0.1 \\
      Max grad norm & 0.1 \\
      \midrule
      \rowcolor{gray!25} \multicolumn{2}{l}{\textbf{Generation Parameters}} \\
      Max prompt length & 4000 \\
      Max answer length & 3000 \\
      Temperature & 0.9 \\
      Top-k & 50 \\
      \bottomrule
    \end{tabular}
  }
  \caption{Configuration parameters used for GRPO training.}
  \label{tab:grpo_training_config}
\end{table}

We train each model using DiverseVul train split, so we consider the DiverseVul test split as \textit{in distribution} data and both CleanVul and BigVul as \textit{out of distribution}.
Table~\ref{tab:grpo_training_config} shows the GRPO training settings. To avoid policy collapse~\cite{policy-collapse}, we chose models at steps with peak or near-peak rewards. Please note that training prompt matches the one used in zero-shot reasoning system-prompt settings (second block of Figure~\ref{fig:prompts_comparison}), so any changes in performance come only from GRPO training, not prompt differences.

\begin{table*}[htbp]
\centering
\setlength{\tabcolsep}{6pt}
\renewcommand{\arraystretch}{1.1}
\rowcolors{3}{gray!10}{white}
\resizebox{0.90\linewidth}{!}{%
\begin{tabular}{ll|ccc|ccc|ccc}
\toprule
\textbf{Category} & \textbf{Metric}
& \multicolumn{3}{c|}{\textbf{Qwen 2.5}}
& \multicolumn{3}{c|}{\textbf{LLaMA 3B}}
& \multicolumn{3}{c}{\textbf{LLaMA 8B}} \\
& & \textbf{DiverseVul} & \textbf{CleanVul} & \textbf{BigVul}
& \textbf{DiverseVul} & \textbf{CleanVul} & \textbf{BigVul}
& \textbf{DiverseVul} & \textbf{CleanVul} & \textbf{BigVul} \\
\midrule
\rowcolor{gray!20}\multicolumn{11}{l}{\textbf{Not Vulnerable}} \\
& Precision
& \cellcolor{green!15}\textbf{0.70} (0.53)
& \cellcolor{green!15}\textbf{0.58} (0.56)
& \cellcolor{green!15}\textbf{0.63} (0.56)
& \cellcolor{green!15}\textbf{0.55} (0.46)
& \cellcolor{green!15}\textbf{0.58} (0.53)
& \cellcolor{green!15}\textbf{0.56} (0.52)
& \cellcolor{green!15}\textbf{0.58} (0.44)
& 0.70 (\textbf{1.00})
& \cellcolor{green!15}\textbf{0.64} (0.51) \\[0.5ex]
& Recall
& \cellcolor{green!15}\textbf{0.61} (0.55)
& 0.54 (\textbf{0.65})
& 0.69 (\textbf{0.91})
& \cellcolor{green!15}\textbf{0.54} (0.39)
& \cellcolor{green!15}\textbf{0.69} (0.47)
& \cellcolor{green!15}\textbf{0.64} (0.53)
& \cellcolor{green!15}\textbf{0.42} (0.26)
& \cellcolor{green!15}\textbf{0.42} (0.00)
& 0.59 (\textbf{0.96}) \\[0.2ex]
& F\textsubscript{1}
& \cellcolor{green!15}\textbf{0.66} (0.54)
& 0.56 (\textbf{0.60})
& 0.66 (\textbf{0.69})
& \cellcolor{green!15}\textbf{0.54} (0.42)
& \cellcolor{green!15}\textbf{0.63} (0.50)
& \cellcolor{green!15}\textbf{0.59} (0.52)
& \cellcolor{green!15}\textbf{0.53} (0.33)
& \cellcolor{green!15}\textbf{0.50} (0.00)
& 0.61 (\textbf{0.67}) \\
\hline
& Support &  2952 & 6330 & 1134 &  2952 & 6330 & 1134 &  2952 & 6330 & 1134 \\
\midrule
\rowcolor{gray!20}\multicolumn{11}{l}{\textbf{Vulnerable}} \\
& Precision
& \cellcolor{green!15}\textbf{0.65} (0.52)
& 0.57 (\textbf{0.59})
& 0.65 (\textbf{0.75})
& \cellcolor{green!15}\textbf{0.52} (0.45)
& \cellcolor{green!15}\textbf{0.62} (0.52)
& \cellcolor{green!15}\textbf{0.55} (0.50)
& \cellcolor{green!15}\textbf{0.58} (0.45)
& \cellcolor{green!15}\textbf{0.56} (0.50)
& \cellcolor{green!15}\textbf{0.61} (0.55) \\[0.2ex]
& Recall
& \cellcolor{green!15}\textbf{0.73} (0.50)
& \cellcolor{green!15}\textbf{0.60} (0.49)
& \cellcolor{green!15}\textbf{0.59} (0.27)
& \cellcolor{green!15}\textbf{0.53} (0.52)
& 0.50 (\textbf{0.58})
& \cellcolor{green!15}\textbf{0.90} (0.50)
& \cellcolor{green!15}\textbf{0.82} (0.65)
& 0.73 (\textbf{1.00})
& \cellcolor{green!15}\textbf{0.91} (0.06) \\[0.2ex]
& F\textsubscript{1}
& \cellcolor{green!15}\textbf{0.69} (0.51)
& \cellcolor{green!15}\textbf{0.58} (0.53)
& \cellcolor{green!15}\textbf{0.62} (0.40)
& \cellcolor{green!15}\textbf{0.52} (0.48)
& \cellcolor{green!15}\textbf{0.56} (0.55)
& \cellcolor{green!15}\textbf{0.64} (0.50)
& \cellcolor{green!15}\textbf{0.68} (0.54)
& 0.63 (\textbf{0.67})
& \cellcolor{green!15}\textbf{0.65} (0.10) \\
\hline
& Support &  2952 & 6330 & 1134 &  2952 & 6330 & 1134 &  2952 & 6330 & 1134 \\
\midrule
\rowcolor{gray!20}\multicolumn{11}{l}{\textbf{Overall}} \\
& Accuracy
& \cellcolor{green!15}\textbf{0.67} (0.52)
& \cellcolor{green!15}\textbf{0.57} (\textbf{0.57})
& \cellcolor{green!15}\textbf{0.64} (0.59)
& \cellcolor{green!15}\textbf{0.58} (0.45)
& \cellcolor{green!15}\textbf{0.60} (0.53)
& \cellcolor{green!15}\textbf{0.55} (0.51)
& \cellcolor{green!15}\textbf{0.62} (0.45)
& \cellcolor{green!15}\textbf{0.58} (0.50)
& \cellcolor{green!15}\textbf{0.62} (0.51) \\[0.2ex]
& Macro Precision
& \cellcolor{green!15}\textbf{0.68} (0.52)
& \cellcolor{green!15}\textbf{0.57} (\textbf{0.57})
& 0.64 (\textbf{0.65})
& \cellcolor{green!15}\textbf{0.65} (0.45)
&  \cellcolor{green!15}\textbf{0.60} (0.53)
& \cellcolor{green!15}\textbf{0.55} (0.51)
& \cellcolor{green!15}\textbf{0.64} (0.45)
& 0.58 (\textbf{0.75})
& \cellcolor{green!15}\textbf{0.62 }(0.53) \\[0.2ex]
& Macro Recall
& \cellcolor{green!15}\textbf{0.67} (0.52)
& \cellcolor{green!15}\textbf{0.57} (\textbf{0.57})
& \cellcolor{green!15}\textbf{0.64} (0.59)
& \cellcolor{green!15}\textbf{0.68} (0.46)
& \cellcolor{green!15}\textbf{0.60} (0.53)
& \cellcolor{green!15}\textbf{0.55} (0.51)
& \cellcolor{green!15}\textbf{0.62} (0.45)
& \cellcolor{green!15}\textbf{0.58} (0.50)
& \cellcolor{green!15}\textbf{0.62} (0.51) \\[0.2ex]
& Macro F\textsubscript{1}
& \cellcolor{green!15}\textbf{0.67} (0.52)
& \cellcolor{green!15}\textbf{0.57} (\textbf{0.57})
& \cellcolor{green!15}\textbf{0.64} (0.54)
& \cellcolor{green!15}\textbf{0.53} (0.45)
& \cellcolor{green!15}\textbf{0.59} (0.53)
& \cellcolor{green!15}\textbf{0.55} (0.51)
& \cellcolor{green!15}\textbf{0.60} (0.43)
& \cellcolor{green!15}\textbf{0.57} (0.33)
& \cellcolor{green!15}\textbf{0.62} (0.38) \\[0.2ex]
& Weighted Precision
& \cellcolor{green!15}\textbf{0.68} (0.52)
& \cellcolor{green!15}\textbf{0.57} (\textbf{0.57})
& 0.64 (\textbf{0.65})
& \cellcolor{green!15}\textbf{0.65} (0.45)
& \cellcolor{green!15}\textbf{0.60} (0.53)
& \cellcolor{green!15}\textbf{0.55} (0.51)
& \cellcolor{green!15}\textbf{0.64} (0.45)
& 0.58 (\textbf{0.75})
& 0.62 (\textbf{0.75}) \\[0.2ex]
& Weighted Recall
& \cellcolor{green!15}\textbf{0.67} (0.52)
& \cellcolor{green!15}\textbf{0.57} (\textbf{0.57})
& \cellcolor{green!15}\textbf{0.64} (0.59)
& \cellcolor{green!15}\textbf{0.58} (0.49)
& \cellcolor{green!15}\textbf{0.60} (0.53)
& \cellcolor{green!15}\textbf{0.55} (0.51)
& \cellcolor{green!15}\textbf{0.62} (0.45)
& \cellcolor{green!15}\textbf{0.58} (0.45)
& \cellcolor{green!15}\textbf{0.62} (0.51) \\[0.2ex]
& Weighted F\textsubscript{1}
& \cellcolor{green!15}\textbf{0.67} (0.52)
& \cellcolor{green!15}\textbf{0.57} (\textbf{0.57})
& \cellcolor{green!15}\textbf{0.64} (0.55)
& \cellcolor{green!15}\textbf{0.53} (0.45)
& \cellcolor{green!15}\textbf{0.59} (0.53)
& \cellcolor{green!15}\textbf{0.55} (0.51)
& \cellcolor{green!15}\textbf{0.60} (0.43)
& \cellcolor{green!15}\textbf{0.57} (0.33)
& \cellcolor{green!15}\textbf{0.62} (0.39) \\
\hline
& Support &  5904 & 12660 & 2268 &  5904 & 12660 & 2268 &  5904 & 12660 & 2268 \\
\bottomrule
\end{tabular}}
\caption{GRPO results across all datasets and models and the best between NR and R in the brackets.}
\label{tab:grpo_all_datasets}
\end{table*}

Table~\ref{tab:grpo_all_datasets} reports GRPO results beside the strongest baseline (the better of the reasoning and no-Reasoning variants), for every combination of model and dataset.  Values outside the brackets refer to GRPO, whereas bracketed numbers correspond to the baseline; green cells highlight a GRPO advantage.
% GRPO outperforms the baseline in roughly three quarters of the entries and, in the few remaining cases, trails by no more than 0.02.  
GRPO consistently achieves superior performance compared to the baseline across all evaluations. On the \textit{in distribution} dataset (DiverseVul) every metric improves for all three models, confirming that the method learns robust decision boundaries without overfitting.  On CleanVul (\textit{out of distribution}) the two approaches are practically tied for Qwen 2.5, while for LLaMA 3B and 8B GRPO removes the tendency to over-predict the vulnerable class and leads to a more balanced precision–recall trade-off.  The largest gains appear on BigVul (\textit{out of distribution}): recall rises sharply, precision remains stable, and models such as Qwen 2.5 and LLaMA 8B shift from a conservative \textit{predict-safe} bias. to a more correct classification of the two classes.

Looking at aggregate metrics, accuracy increases by 4–17 percentage points for all models (except for Cleanvul in Qwen 2.5 whose accuracy remains stable). Macro and weighted F\textsubscript{1} move in lock-step with accuracy, indicating that improvements benefit both classes rather than a single dominant one.  

%In summary, the results confirm that GRPO generalizes effectively across datasets, even in out-of-distribution scenarios. Additionally, GRPO mitigates unbalanced classification tendencies and demonstrates scalability, maintaining or enhancing its effectiveness when applied to larger models such as LLaMA 3B and 8B.

To summarize the results achieved in this section, we conclude the following:

\begin{tcolorbox}[colback=black!1!white,colframe=black!1!black]
    \begin{minipage}{\columnwidth}
\RQ{RQ2} \textit{Can we train the model to use its own reasoning to identify vulnerabilities?} 

\RQ{A2} \textit{The proposed GRPO formulation enables models to harness their own reasoning to identify vulnerabilities, consistently outperforming baselines across both in distribution and out of distribution datasets.}
\end{minipage}
\end{tcolorbox}%

% \begin{tcolorbox}[title=Q2\label{q2}, colback=red!3!white, colframe=red!30!black]
% Can we train the model to leverage its own \textbf{reasoning} to identify vulnerabilities?
% \end{tcolorbox}

% % E nella risposta:
% \begin{tcolorbox}[title=A2, colback=green!3!white, colframe=green!30!black]
% \textbf{\textit{Yes. GRPO enables models to harness their own reasoning to identify vulnerabilities, consistently outperforming baselines across both in-distribution and out-of-distribution datasets.}}
% \end{tcolorbox}

\section{Comparing GRPO and SFT}
\label{sec:GRPOvsSFT}

Table~\ref{tab:grpo_all_datasets} shows that models trained with GRPO outperform the best baseline (either \textit{Reasoning} or \textit{No-Reasoning}) across all model and dataset combinations. However, it is important to assess whether this improvement comes from the use of reasoning or simply from finetuning the model on the vulnerability detection task.
To investigate this, we finetuned the models using standard SFT, following common practice in recent LLM-based vulnerability detection research~\cite{Harnessing_LLMs, LLMAO}. In SFT, the model is trained to directly predict the final verdict without generating any reasoning. We then compared the performance of GRPO and SFT to understand the real impact of reasoning.

\begin{table}[ht]
  \centering
  \scriptsize
  \renewcommand{\arraystretch}{0.4}  % Reduce row height

  \rowcolors{3}{gray!10}{white}
  \resizebox{0.85\linewidth}{!}{%
    \begin{tabular}{l|l}
      \toprule
      \rowcolor{gray!25} \multicolumn{2}{l}{\textbf{Training Setup}} \\
      N. train epochs & 2 \\
      Batch size & 16 \\
      Gradient accumulation steps & 2 \\
      Save strategy & epoch \\
      Logging steps & 1 \\
      \midrule
      \rowcolor{gray!25} \multicolumn{2}{l}{\textbf{Optimization Parameters}} \\
      Learning rate & 3e-4 \\
      Weight decay & 0.01 \\
      Adam beta1 & 0.9 \\
      Adam beta2 & 0.99 \\
      Optim & adamw\_8bit \\
      LR scheduler type & linear \\
      Warmup steps & 10 \\
      Max grad norm & 0.1 \\
      \midrule
      \rowcolor{gray!25} \multicolumn{2}{l}{\textbf{Generation Parameters}} \\
      Max prompt length & 4000 \\
      Max answer length & 1000 \\
      Temperature & 0.9 \\
      Top-k & 50 \\
      \bottomrule
    \end{tabular}
  }
  \caption{Configuration parameters used for SFT training.}
  \label{tab:sft_training_config}
\end{table}

Even in this case, we finetuned the models using the training split of DiverseVul, so we can still consider the test split of DiverseVul as \textit{in distribution} and BigVul and CleanVul as \textit{out of distribution}. The parameter used to train SFT are shown in table~\ref{tab:sft_training_config}.
All models finetuned using SFT were trained with the no-reasoning system prompt in Figure~\ref{fig:prompts_comparison}, used in the zero-shot setting to isolate the effect of the finetuning strategy from differences in instruction formatting. 

\begin{table}[htbp]
    % Flush left instead of centered
    \raggedright
    % Slightly larger font now that we save horizontal space
    \small
    % Tighten inter‑column spacing
    \setlength{\tabcolsep}{3pt}
    % Align on decimal separator
    \sisetup{round-mode=places, round-precision=2, table-format=1.2}
    % Alternate row shading for readability
    \rowcolors{3}{gray!8}{white}

    % Scale to \linewidth while preserving left alignment
    \resizebox{\linewidth}{!}{%
        \begin{tabular}{@{}ll|*{3}{S}|*{3}{S}|*{3}{S}@{}}
            \toprule
            \textbf{Category} & \textbf{Metric}
                & \multicolumn{3}{c|}{\textbf{Qwen 2.5}}
                & \multicolumn{3}{c|}{\textbf{LLaMA 3B}}
                & \multicolumn{3}{c}{\textbf{LLaMA 8B}} \\
            & & \textbf{Div.} & \textbf{Cln.} & \textbf{Big}
              & \textbf{Div.} & \textbf{Cln.} & \textbf{Big}
              & \textbf{Div.} & \textbf{Cln.} & \textbf{Big} \\
            \midrule

            \rowcolor{gray!20} \multicolumn{11}{l}{\textbf{Not Vulnerable}} \\
            & \textbf{Precision}          & 0.48 & 0.45 & 0.55 & 0.76 & 0.55 & 0.55 & 0.88 & 0.52 & 0.61 \\
            & \textbf{Recall}             & 0.07 & 0.08 & 0.13 & 0.24 & 0.10 & 0.13 & 0.43 & 0.09 & 0.14 \\
            & \textbf{F\textsubscript{1}} & 0.12 & 0.13 & 0.22 & 0.37 & 0.17 & 0.21 & 0.58 & 0.16 & 0.23 \\

            \midrule
            \rowcolor{gray!20} \multicolumn{11}{l}{\textbf{Vulnerable}} \\
            & \textbf{Precision}          & 0.50 & 0.50 & 0.51 & 0.52 & 0.51 & 0.50 & 0.62 & 0.50 & 0.51 \\
            & \textbf{Recall}             & 0.93 & 0.60 & 0.59 & 0.53 & 0.51 & 0.45 & 0.42 & 0.73 & 0.66 \\
            & \textbf{F\textsubscript{1}} & 0.65 & 0.58 & 0.62 & 0.52 & 0.56 & 0.50 & 0.53 & 0.63 & 0.63 \\

            \midrule
            \rowcolor{gray!20} \multicolumn{11}{l}{\textbf{Overall}} \\
            & \textbf{Accuracy}           & 0.50 & 0.49 & 0.51 & 0.53 & 0.51 & 0.51 & 0.62 & 0.50 & 0.52 \\
            & \textbf{Macro Precision}    & 0.49 & 0.47 & 0.53 & 0.53 & 0.53 & 0.53 & 0.64 & 0.51 & 0.56 \\
            & \textbf{Macro Recall}       & 0.50 & 0.49 & 0.51 & 0.58 & 0.51 & 0.51 & 0.62 & 0.50 & 0.53 \\
            & \textbf{Macro F\textsubscript{1}} & 0.38 & 0.39 & 0.43 & 0.52 & 0.41 & 0.42 & 0.60 & 0.40 & 0.44 \\
            & \textbf{Weighted Precision} & 0.49 & 0.47 & 0.53 & 0.53 & 0.53 & 0.53 & 0.64 & 0.51 & 0.56 \\
            & \textbf{Weighted Recall}    & 0.50 & 0.49 & 0.51 & 0.53 & 0.51 & 0.51 & 0.62 & 0.50 & 0.52 \\
            & \textbf{Weighted F\textsubscript{1}} & 0.38 & 0.39 & 0.43 & 0.52 & 0.41 & 0.42 & 0.60 & 0.40 & 0.44 \\
            \bottomrule
        \end{tabular}%
    }

    \caption{SFT results (rounded to two decimals). Div.=DiverseVul, Cln.=CleanVul, Big=BigVul.}
    \label{tab:sft_results_all}
\end{table}

\subsection{Comparative Analysis across datasets}
Table~\ref{tab:sft_results_all} shows that SFT leaves the models heavily biased.  For Qwen 2.5 and LLaMA 3B, the \textit{Not Vulnerable} class is almost ignored: recall never exceeds 0.24 and the class F\textsubscript{1} peaks at 0.37.  LLaMA 8B does a little better on the in distribution split (\textit{DiverseVul}, recall 0.43, F\textsubscript{1} 0.58) but falls back to recall 0.09 and 0.14 on the two out of distribution sets, CleanVul and BigVul respectively.  
The \textit{Vulnerable} class shows the opposite pattern: recall is high, up to 0.93 for Qwen 2.5 on \textit{DiverseVul}, yet precision stays near 0.50.  
Because of this imbalance, macro F\textsubscript{1} ranges only from 0.38 to 0.60 and drops sharply whenever the test data deviate from the training distribution (for example, LLaMA 8B falls from 0.60 on \textit{DiverseVul} to 0.40 on \textit{CleanVul}).

Comparing the results of SFT in Table~\ref{tab:sft_results_all} and the results of GRPO in Table~\ref{tab:comparison-R-NR}, the latter leads to clear improvements over SFT across all models and datasets. Macro\,F\textsubscript{1} increases by 1-29 points, reaching up to 0.67 on Qwen 2.5 (\textit{DiverseVul}), 0.59 on LLaMA 3B (\textit{CleanVul}), and 0.62 on LLaMA 8B (\textit{BigVul}. Accuracy also improves, with all models scoring between 0.58 and 0.67. Most of the gain comes from the \textit{Not Vulnerable} class. For example, recall goes from 0.07 to 0.61 on Qwen 2.5 (\textit{DiverseVul}) and from 0.09 to 0.42 on LLaMA 8B (\textit{CleanVul}). Precision stays stable, so the F\textsubscript{1} score improves significantly. The \textit{Vulnerable} class also maintains high recall and sometimes gains precision.
Since these improvements also appear on \textit{out of distribution} datasets, they are likely due to the reasoning encouraged by GRPO rather than simple finetuning.

\subsection{Comparative analysis across programming languages}
In this part, we evaluate the performance of the two training strategies in detecting vulnerabilities without relying on language-specific patterns.  
To this end, we test models trained on \textit{CleanVul} using code examples written in \textit{JavaScript}, \textit{Python}, and \textit{Java}.  
These programming languages were not seen during training on \textit{DiverseVul}, which contains only C code.

Table~\ref{tab:grpo_sft_languages} shows that GRPO improves performance across all languages and models. For example, macro F\textsubscript{1} on Java increases from 0.25 to 0.46 with Qwen 2.5, and from 0.34 to 0.58 on Python. LLaMA 3B and 8B also show consistent gains, especially on Python and Java. Accuracy and weighted F\textsubscript{1} follow the same trend.  
Most improvements come from better handling of the \textit{Not Vulnerable} class: GRPO raises its F\textsubscript{1} while maintaining high precision. The \textit{Vulnerable} class keeps good precision, resulting in better balance overall.
Since these gains appear even on languages not used during training, the results suggest also here that GRPO reasoning step helps models generalize better across programming languages.

%In summary, GRPO consistently outperforms SFT across all evaluation axes: it improves macro and weighted F\textsubscript{1}, enhances recall on the underrepresented \textit{Not Vulnerable} class, and generalises better to unseen data. These gains extend to doubly out-of-distribution settings—where models are tested on languages not seen during training—confirming GRPO's ability to capture transferable vulnerability patterns. All improvements come without architectural changes or inference overhead, making GRPO a robust and generalisable alternative to SFT for real-world vulnerability detection.

\begin{figure*}[ht]  % se usi una classe a due colonne vedi nota sotto
    \centering
    % --------  (a) DiverseVul  ----------
    \begin{subfigure}{0.45\linewidth}
        \includegraphics[width=\linewidth]{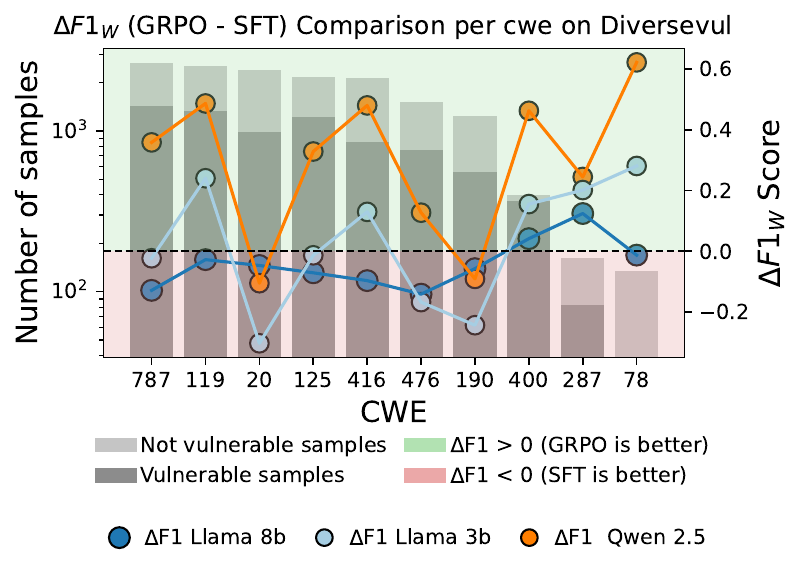}
        %\caption{DiverseVul}%
        %\label{fig:CWE_diversevul_dataset}
    \end{subfigure}%   ← niente riga vuota né \par qui!
    \hfill                     % spaziatore orizzontale
    % --------  (b) BigVul  --------------
    \begin{subfigure}{0.45\linewidth}
        \includegraphics[width=\linewidth]{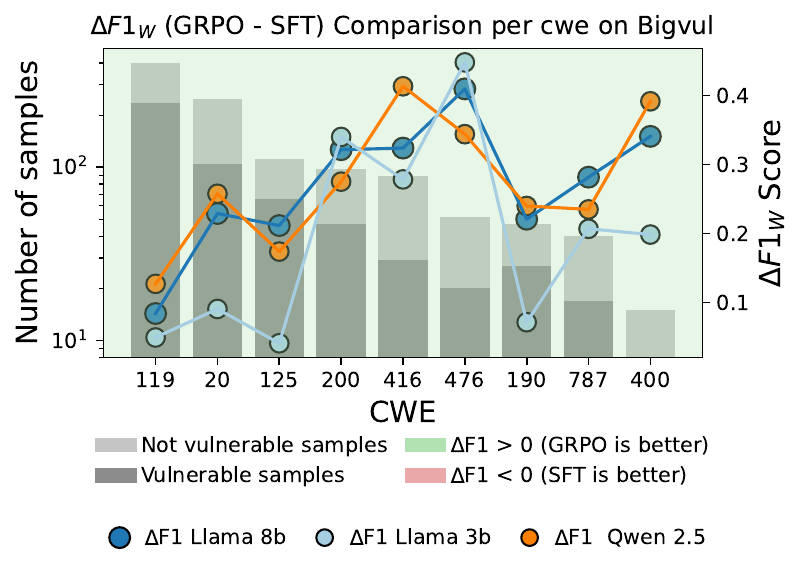}
        %\caption{BigVul}%
        %\label{fig:CWE_bigvul_dataset}
    \end{subfigure}

    \caption{Comparison of CWE performance for GRPO-SFT across three models. The 15 CWEs shown are part of MITRE’s Top 25.}
    \label{fig:CWE_models_comparison}
\end{figure*}

\subsection{Comparative analysis across CWEs}
To see how GRPO and SFT affect specific weakness types, we compare their performance on ten \textit{Common Weakness Enumeration} (CWE) taken from the \textit{2024 CWE Top 25 Most Dangerous Software Weaknesses}~\footnote{\url{https://cwe.mitre.org/top25/archive/2024/2024_cwe_top25.html#top25list}}.  
As shown in Fig.~\ref{fig:CWE_models_comparison}, These CWEs appear in both \textit{DiverseVul} and \textit{BigVul}. For each CWE, we calculate the change in weighted \(F_{1}\) score (\(\Delta F_{1}^{w}\)) by subtracting SFT performance from GRPO performance.

On \textit{DiverseVul} GRPO strengthens smaller models. Qwen 2.5 shows significant improvements in most CWEs, especially in \textit{CWE-78}, \textit{CWE-119} and \textit{CWE-416}. LLaMA 3B shows mixed results. It improves on some CWEs but drops on others such as \textit{CWE-20} and \textit{CWE-190}, suggesting that GRPO sometimes over-adjusts when the base model is already performing well. LLaMA 8B shows little to no improvement of \textit{DiverseVul} and even some small decreases (e.g. \textit{CWE-787}, \textit{CWE-476}). This suggests that the SFT allows the model to memorize patterns during training, while the abstract reasoning of GRPO is not able to achieve this level of memorization.

On \textit{BigVul}, GRPO consistently outperforms SFT across all CWEs and model sizes. Among the evaluated models, Qwen 2.5 exhibits the most significant overall improvement, highlighting the particular advantage of GRPO for smaller models. Similarly, LLaMA 8B follows this positive trend, while LLaMA 3B achieves its most notable gains on \textit{CWE-476}.

%Qwen 2.5 also improves across the board, confirming that GRPO benefits smaller models, while LLaMA 8B follows the same positive trend.

% Overall, GRPO acts as a tuning method that adapts to model size. It brings strong, general improvements for smaller models, more targeted effects for mid sized ones, and helps models shift from memorizing to reasoning. These patterns hold across both datasets, showing that GRPO’s effects are due to how it interacts with model capacity and vulnerability type, not just dataset quirks.

% Although these results are not on real-world data, the consistent improvement of GRPO on out-of-distribution datasets (BigVul and Cleanvul) suggests that training with GRPO could also help in such cases. We leave this for future work.

To summarize the results in this comparative analysis between GRPO and SFT, we conclude the following:  

\begin{tcolorbox}[colback=black!1!white,colframe=black!1!black]
    \begin{minipage}{\columnwidth}
\RQ{RQ3} \textit{In what ways does GRPO differ from (and potentially improve upon) traditional supervised finetuning for code vulnerability detection?}

\RQ{A3} \textit{GRPO consistently improves overall performance, enhances recall on non vulnerable code, and maintains robustness under distribution shift, including on unseen programming languages.}
\end{minipage}
\end{tcolorbox}%

\begin{table*}[htbp]
\centering
\rowcolors{3}{gray!10}{white}
\begin{adjustbox}{max width=\linewidth}
\small
\setlength{\tabcolsep}{4pt}
\resizebox{0.85\textwidth}{!}{%
\begin{tabular}{ll|ccc|ccc|ccc}
\toprule
\textbf{Category} & \textbf{Metrics}
& \multicolumn{3}{c|}{\textbf{Qwen 2.5}}
& \multicolumn{3}{c|}{\textbf{LLaMA 3B}}
& \multicolumn{3}{c}{\textbf{LLaMA 8B}} \\
& & JS & Py & Ja
  & JS & Py & Ja
  & JS & Py & Ja \\
\midrule
\rowcolor{gray!25} \multicolumn{11}{l}{\textbf{Not Vulnerable}} \\
 & Precision
   & \cellcolor{red!20}0.61 (\textbf{0.76})
   & \cellcolor{green!20}\textbf{0.70} (0.49)
   & \cellcolor{red!20}0.69 (\textbf{0.85})
   & \cellcolor{green!20}\textbf{0.62} (0.59)
   & \cellcolor{red!20}0.69 (\textbf{0.71})
   & \cellcolor{green!20}\textbf{0.74} (0.63)
   & \cellcolor{green!20}\textbf{0.62} (0.55)
   & \cellcolor{green!20}\textbf{0.72} (0.68)
   & \cellcolor{green!20}\textbf{0.72} (0.69) \\

 & Recall
   & \cellcolor{green!20}\textbf{0.77} (0.20)
   & \cellcolor{green!20}\textbf{0.56} (0.09)
   & \cellcolor{green!20}\textbf{0.53} (0.02)
   & \cellcolor{green!20}\textbf{0.65} (0.09)
   & \cellcolor{green!20}\textbf{0.74} (0.25)
   & \cellcolor{green!20}\textbf{0.72} (0.04)
   & \cellcolor{green!20}\textbf{0.49} (0.22)
   & \cellcolor{green!20}\textbf{0.44} (0.10)
   & \cellcolor{green!20}\textbf{0.46} (0.03) \\

 & F\textsubscript{1}
   & \cellcolor{green!20}\textbf{0.68} (0.32)
   & \cellcolor{green!20}\textbf{0.62} (0.15)
   & \cellcolor{green!20}\textbf{0.60} (0.05)
   & \cellcolor{green!20}\textbf{0.63} (0.16)
   & \cellcolor{green!20}\textbf{0.71} (0.37)
   & \cellcolor{green!20}\textbf{0.73} (0.07)
   & \cellcolor{green!20}\textbf{0.55} (0.32)
   & \cellcolor{green!20}\textbf{0.55} (0.17)
   & \cellcolor{green!20}\textbf{0.56} (0.06) \\

\hline
 & Support & 1229 & 1495 & 3040 & 1229 & 1495 & 3040 & 1229 & 1495 & 3040 \\

\midrule
\rowcolor{gray!25} \multicolumn{11}{l}{\textbf{Vulnerable}} \\
 & Precision
   & \cellcolor{green!20}\textbf{0.46} (0.44)
   & \cellcolor{green!20}\textbf{0.47} (0.37)
   & \cellcolor{red!20}0.27 (\textbf{0.29})
   & \cellcolor{green!20}\textbf{0.45} (0.41)
   & \cellcolor{green!20}\textbf{0.53} (0.41)
   & \cellcolor{green!20}\textbf{0.37} (0.29)
   & \cellcolor{green!20}\textbf{0.43} (0.40)
   & \cellcolor{green!20}\textbf{0.45} (0.40)
   & \cellcolor{green!20}\textbf{0.30} (0.29) \\

 & Recall
   & \cellcolor{red!20}0.29 (\textbf{0.91})
   & \cellcolor{red!20}0.63 (\textbf{0.85})
   & \cellcolor{red!20}0.42 (\textbf{0.99})
   & \cellcolor{red!20}0.42 (\textbf{0.91})
   & \cellcolor{red!20}0.47 (\textbf{0.84})
   & \cellcolor{red!20}0.40 (\textbf{0.95})
   & \cellcolor{red!20}0.56 (\textbf{0.74})
   & \cellcolor{red!20}0.73 (\textbf{0.93})
   & \cellcolor{red!20}0.57 (\textbf{0.96}) \\

 & F\textsubscript{1}
   & \cellcolor{red!20}0.36 (\textbf{0.59})
   & \cellcolor{green!20}\textbf{0.54} (0.52)
   & \cellcolor{red!20}0.33 (\textbf{0.45})
   & \cellcolor{red!20}0.43 (\textbf{0.56})
   & \cellcolor{red!20}0.50 (\textbf{0.55})
   & \cellcolor{red!20}0.38 (\textbf{0.44})
   & \cellcolor{red!20}0.49 (\textbf{0.52})
   & \cellcolor{green!20}\textbf{0.56} (0.55)
   & \cellcolor{red!20}0.40 (\textbf{0.45}) \\

\hline
 & Support & 976 & 971 & 1244 & 976 & 971 & 1244 & 976 & 971 & 1244\\

\midrule
\rowcolor{gray!25} \multicolumn{11}{l}{\textbf{Overall}} \\
 & Accuracy
   & \cellcolor{green!20}\textbf{0.57} (0.49)
   & \cellcolor{green!20}\textbf{0.58} (0.39)
   & \cellcolor{green!20}\textbf{0.50} (0.31)
   & \cellcolor{green!20}\textbf{0.56} (0.42)
   & \cellcolor{green!20}\textbf{0.64} (0.48)
   & \cellcolor{green!20}\textbf{0.62} (0.30)
   & \cellcolor{green!20}\textbf{0.52} (0.43)
   & \cellcolor{green!20}\textbf{0.55} (0.42)
   & \cellcolor{green!20}\textbf{0.49} (0.30) \\

 & Macro Precision
   & \cellcolor{red!20}0.54 (\textbf{0.60})
   & \cellcolor{green!20}\textbf{0.59} (0.43)
   & \cellcolor{red!20}0.48 (\textbf{0.57})
   & \cellcolor{green!20}\textbf{0.53} (0.50)
   & \cellcolor{green!20}\textbf{0.61} (0.56)
   & \cellcolor{green!20}\textbf{0.56} (0.46)
   & \cellcolor{green!20}\textbf{0.53} (0.47)
   & \cellcolor{green!20}\textbf{0.59} (0.54)
   & \cellcolor{green!20}\textbf{0.51} (0.49) \\

 & Macro Recall
   & \cellcolor{red!20}0.53 (\textbf{0.56})
   & \cellcolor{green!20}\textbf{0.59} (0.47)
   & \cellcolor{red!20}0.48 (\textbf{0.51})
   & \cellcolor{green!20}\textbf{0.53} (0.50)
   & \cellcolor{green!20}\textbf{0.61} (0.54)
   & \cellcolor{green!20}\textbf{0.56} (0.49)
   & \cellcolor{green!20}\textbf{0.53} (0.48)
   & \cellcolor{green!20}\textbf{0.58} (0.51)
   & \cellcolor{green!20}\textbf{0.51} (0.50) \\

 & Macro F\textsubscript{1}
   & \cellcolor{green!20}\textbf{0.52} (0.46)
   & \cellcolor{green!20}\textbf{0.58} (0.34)
   & \cellcolor{green!20}\textbf{0.46} (0.25)
   & \cellcolor{green!20}\textbf{0.53} (0.36)
   & \cellcolor{green!20}\textbf{0.61} (0.46)
   & \cellcolor{green!20}\textbf{0.56} (0.25)
   & \cellcolor{green!20}\textbf{0.52} (0.42)
   & \cellcolor{green!20}\textbf{0.55} (0.36)
   & \cellcolor{green!20}\textbf{0.48} (0.25) \\

 & Weighted Precision
   & \cellcolor{red!20}0.55 (\textbf{0.63})
   & \cellcolor{green!20}\textbf{0.61} (0.45)
   & \cellcolor{red!20}0.57 (\textbf{0.69})
   & \cellcolor{green!20}\textbf{0.55} (0.52)
   & \cellcolor{green!20}\textbf{0.63} (0.59)
   & \cellcolor{green!20}\textbf{0.63} (0.53)
   & \cellcolor{green!20}\textbf{0.54} (0.49)
   & \cellcolor{green!20}\textbf{0.61} (0.57)
   & \cellcolor{green!20}\textbf{0.60} (0.57) \\

 & Weighted Recall
   & \cellcolor{green!20}\textbf{0.57} (0.49)
   & \cellcolor{green!20}\textbf{0.58} (0.39)
   & \cellcolor{green!20}\textbf{0.50} (0.31)
   & \cellcolor{green!20}\textbf{0.56} (0.42)
   & \cellcolor{green!20}\textbf{0.64} (0.48)
   & \cellcolor{green!20}\textbf{0.62} (0.30)
   & \cellcolor{green!20}\textbf{0.52} (0.43)
   & \cellcolor{green!20}\textbf{0.55} (0.42)
   & \cellcolor{green!20}\textbf{0.49} (0.30) \\

 & Weighted F\textsubscript{1}
   & \cellcolor{green!20}\textbf{0.55} (0.43)
   & \cellcolor{green!20}\textbf{0.59} (0.29)
   & \cellcolor{green!20}\textbf{0.52} (0.17)
   & \cellcolor{green!20}\textbf{0.55} (0.32)
   & \cellcolor{green!20}\textbf{0.63} (0.44)
   & \cellcolor{green!20}\textbf{0.63} (0.18)
   & \cellcolor{green!20}\textbf{0.52} (0.40)
   & \cellcolor{green!20}\textbf{0.55} (0.32)
   & \cellcolor{green!20}\textbf{0.51} (0.17) \\
\hline
 & Support & 2205 & 2466 & 4284 & 2205 & 2466 & 4284 & 2205 & 2466 & 4284 \\
\bottomrule
\end{tabular}
}
\end{adjustbox}
\caption{GRPO vs. SFT for programming languages extracted from CleanVul dataset different from the progamming language used during training (C)}
\label{tab:grpo_sft_languages}
\end{table*}

\section{Ablation Studies}
\label{sec:ablation}

In this section, we aim to provide ablations studies and evaluate how effectively training with GRPO has enhanced the models reasoning capabilities.

%   \includegraphics[width=\linewidth]
%     {images/New_diversevul_reasoning_lengths_smoothed_multi_with_means.pdf}
%   \caption{Answer length comparison between GRPO and SFT on \textbf{DiverseVul}}
%   \label{fig:length_div}
% \end{figure}

% % Figura 2 – BigVul
% \begin{figure}[htbp]
%   \centering
%   \includegraphics[width=\linewidth]
%     {images/New_BigVul_reasoning_lengths_smoothed_multi_with_means.pdf}
%   \caption{Answer length comparison between GRPO and SFT on \textbf{BigVul}}
%   \label{fig:length_big}
% \end{figure}

\subsection{Linking Model Reasoning to CWE Meaning}
\begin{figure*}[t]
  \centering
  % Riga 1 - LLaMA 8B
  \begin{subfigure}[b]{0.40\textwidth}
    \includegraphics[width=\textwidth]{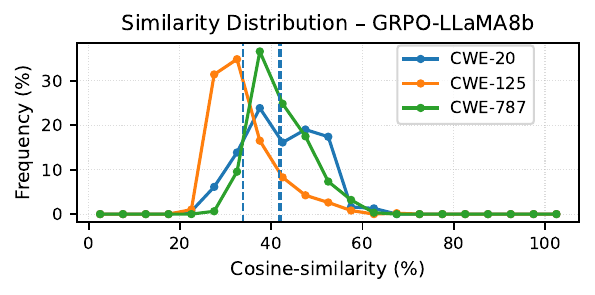}
  \end{subfigure}
  \begin{subfigure}[b]{0.40\textwidth}
    \includegraphics[width=\textwidth]{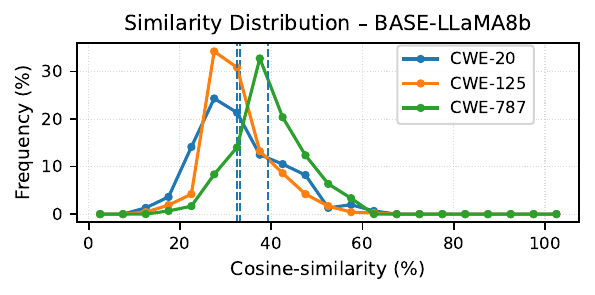}
  \end{subfigure}

  % Riga 2 - LLaMA 3B
  \begin{subfigure}[b]{0.40\textwidth}
    \includegraphics[width=\textwidth]{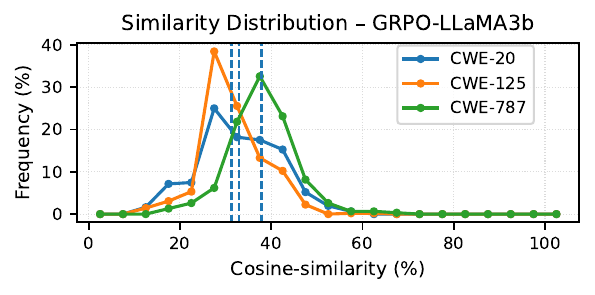}
  \end{subfigure}
  \begin{subfigure}[b]{0.40\textwidth}
    \includegraphics[width=\textwidth]{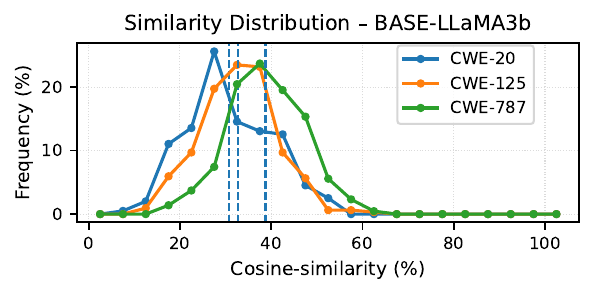}
  \end{subfigure}

  % Riga 3 - Qwen
  \begin{subfigure}[b]{0.40\textwidth}
    \includegraphics[width=\textwidth]{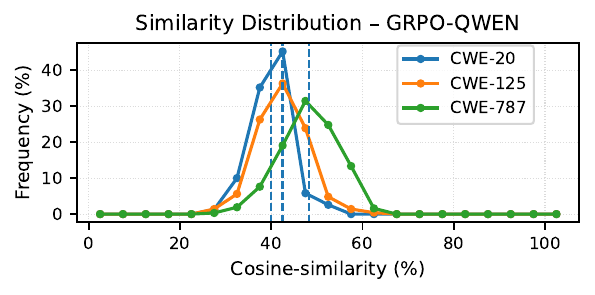}
  \end{subfigure}
  \begin{subfigure}[b]{0.40\textwidth}
    \includegraphics[width=\textwidth]{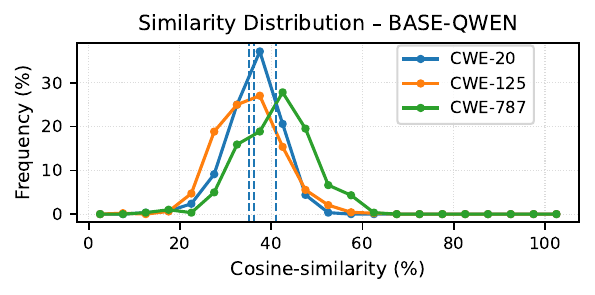}
  \end{subfigure}

  \caption{Cosine-similarity distribution for the 10 CWE categories, comparing the GRPO and BASE models on the three models (LLaMA 8B, LLaMA 3B, Qwen).}
  \label{fig:sim-grpo-vs-BASE-all}
\end{figure*}

We want to know whether the explanation the model writes in \textbf{Step 2} in the System Prompt for \textit{Reasoning Test}(\textit{“Check thoroughly for vulnerabilities ...’’} in Fig.~\ref{fig:prompts_comparison}) actually matches the CWE weakness present in the code. For every test sample we embed that sentence with \texttt{MiniLM-L6-} \texttt{v2} and compare it, via cosine similarity, to the official CWE description taken from the MITRE site\footnote{\url{https://cwe.mitre.org/}}.  Higher similarity means the model is talking about the right vulnerability.  The score distributions for GRPO and SFT are plotted in Fig.~\ref{fig:sim-grpo-vs-BASE-all}.  To see whether the two curves differ in a statistically meaningful way we apply a two-sample Kolmogorov–Smirnov test (KS)~\cite{smirnov1948table} with a 0.01 significance level. Table~\ref{tab:k-s_sim} summarises the outcome for the three most frequent CWEs in the \textit{DiverseVul} split.  Qwen 2.5 and LLaMA 8B show higher similarities for every CWE, and the KS test confirms that GRPO moves the whole distribution, not just the average.  LLaMA 3B improves on CWE-20, stays unchanged on CWE-787, and drops slightly on CWE-125, reflecting its smaller capacity and more limited exposure to security code during pre-training.  In all cases the similarity scores become less spread out, indicating that GRPO makes the model’s explanations more consistently aligned with the correct weakness.

Because the models never saw explicit CWE labels during training, these gains must come from better use of prior knowledge rather than simple memorisation.  GRPO therefore does more than tweak predictions: it pushes the model to reason with concepts that match the real CWE definitions.

\subsection{Concise and Correct: Analyzing Reasoning Length}

\begin{figure}[!ht]
  \centering
  \begin{subfigure}[b]{0.35\textwidth}
    \includegraphics[width=\textwidth]{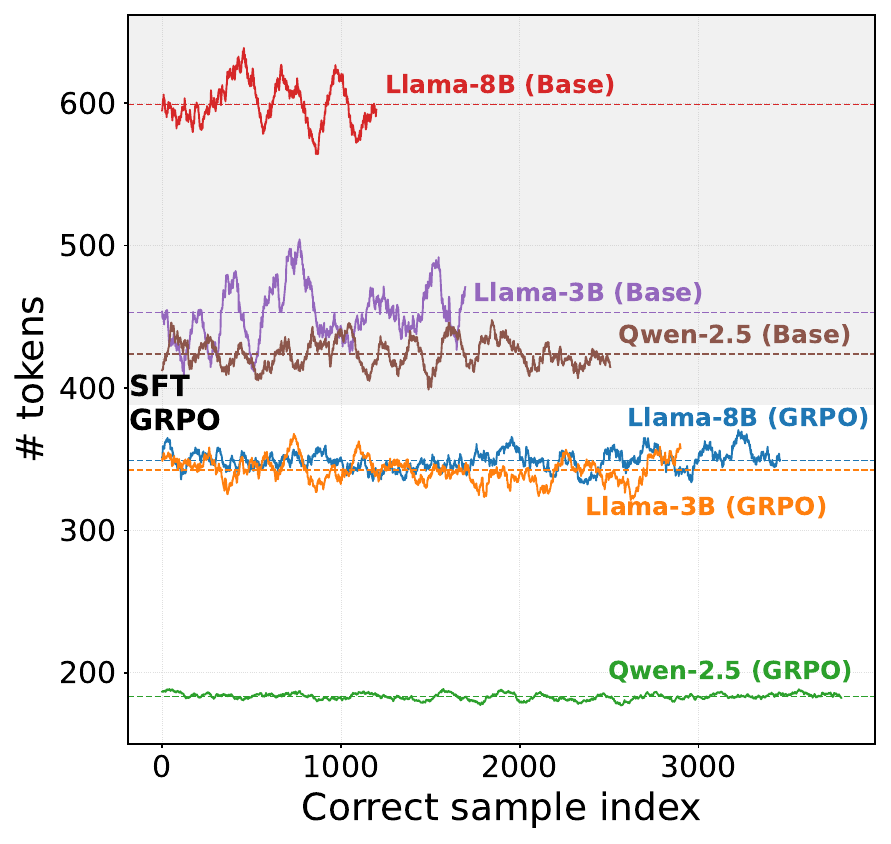}
    \caption{DiverseVul}
    \label{fig:length_div}
  \end{subfigure}
  \begin{subfigure}[b]{0.35\textwidth}
    \includegraphics[width=\textwidth]{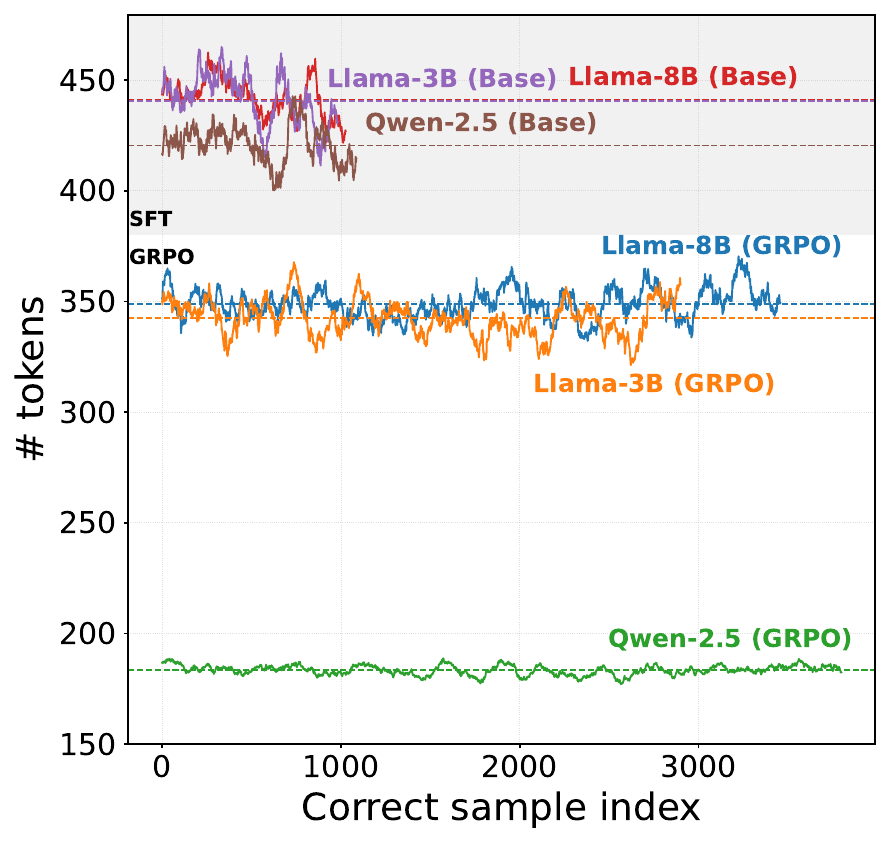}
    \caption{BigVul}
    \label{fig:length_big}
  \end{subfigure}
  \caption{Answer length comparison between GRPO and SFT}
%   \label{fig:length_big}
\end{figure}

Figure~\ref{fig:length_big} shows the number of tokens used in the \verb|<reasoning>| section for each test example where the models trained with GRPO and SFT provided the correct prediction.
The \textbf{x-axis} represents the index of each correct example, while the \textbf{y-axis} indicates the length of the explanation in tokens (we used \texttt{BERT} tokenizer~\footnote{\url{https://huggingface.co/google-bert/bert-base-uncased}} for each model). For both datasets, DiverseVul (top) and BigVul (bottom), we observe a consistent pattern: models finetuned with GRPO produce significantly shorter explanations compared to their baseline versions (SFT). 
On DiverseVul, for example, LLaMA 8B (Base) produces reasonings with an average of around 600 tokens, while the GRPO variant reduces this to around 340 tokens.
A similar trend applies to all other models. As only correct predictions are considered, this reduction in length suggests that GRPO encourages more focused and efficient reasoning, the same technical insight is conveyed in fewer words.
This conciseness is particularly valuable in real-world security operations, where clarity and speed of understanding are critical.
\begin{table}[ht]
  \centering
  \small                         % smaller font
  
  \label{tab:stats_cwe_models}
  \rowcolors{3}{gray!10}{white}  % optional: alternate row colors
  \resizebox{0.95\linewidth}{!}{%
    \begin{tabular}{l|cc|cc|cc}
      \toprule
      \multirow{2}{*}{\textbf{Metric}}
        & \multicolumn{2}{c|}{\textbf{LLaMA 8B}}
        & \multicolumn{2}{c|}{\textbf{LLaMA 3B}}
        & \multicolumn{2}{c}{\textbf{Qwen 2.5}} \\
      \cmidrule(lr){2-3}\cmidrule(lr){4-5}\cmidrule(lr){6-7}
        & BASE & GRPO & BASE & GRPO & BASE & GRPO \\
      \midrule

      %--- CWE-787 block (centered) ---%
      \rowcolor{gray!25} \multicolumn{7}{c}{\textbf{CWE-787}}\\[-0.8ex]
      \textbf{Same Distribution?}
        & \multicolumn{2}{c|}{No}
        & \multicolumn{2}{c|}{Yes}
        & \multicolumn{2}{c}{No} \\
      % \textbf{Gaussian distribution}
      %   & Yes  & No   & Yes & Yes & Yes & Yes \\
      Mean      & 39.37  & \cellcolor{green!20} \textbf{41.69}  & \cellcolor{gray!20}38.75 & \cellcolor{gray!20}37.87 & 41.13 & \cellcolor{green!20} \textbf{48.33} \\
      % Median    & 38.60  & \cellcolor{green!20} \textbf{40.98}  & \cellcolor{gray!20}38.41 & \cellcolor{gray!20}37.53 & 41.65 & \cellcolor{green!20} \textbf{48.02} \\
      Std.\ dev.&  \textbf{7.51}  & \cellcolor{green!20}  6.31  & \cellcolor{gray!20}8.29 & \cellcolor{gray!20}7.00 & \textbf{7.93} & \cellcolor{green!20} 6.28 \\
      % Skewness  &  0.21  &  0.69  & $-$0.02 & 0.31 & $-$0.16 & $-$0.22 \\
      % Kurtosis  &  0.04  &  0.11  & 0.10 & 2.22 & 0.33 & $-$0.28 \\

      %--- CWE-20 block ---%
      \rowcolor{gray!25} \multicolumn{7}{c}{\textbf{CWE-20}}\\[-0.8ex]
      \textbf{Same Distribution?}
        & \multicolumn{2}{c|}{No}
        & \multicolumn{2}{c|}{No}
        & \multicolumn{2}{c}{No} \\
      % \textbf{Gaussian distribution}
      %   & No   & Yes   & Yes & Yes & Yes & No \\
      Mean      & 33.11  & \cellcolor{green!20} \textbf{42.17}  & 30.73 & \cellcolor{green!20} \textbf{32.94} & 36.23 & \cellcolor{green!20} \textbf{39.93} \\
      % Median    & 31.41  & \cellcolor{green!20} \textbf{42.39}  & 29.60 & \cellcolor{green!20} \textbf{32.62} & 36.38 & \cellcolor{green!20} \textbf{40.72} \\
      Std.\ dev.&  \textbf{9.30}  & \cellcolor{green!20} 8.03  & \textbf{9.49 }& \cellcolor{green!20} 8.69 & \textbf{5.66} & \cellcolor{green!20} 4.15 \\
      % Skewness  &  0.66  &  0.01  & 0.20 & 0.02 & $-$0.55 & 0.13 \\
      % Kurtosis  &  0.27  & $-$0.81& $-$0.41 & $-$0.26 & 1.34 & 1.01 \\

      %--- CWE-125 block ---%
      \rowcolor{gray!25} \multicolumn{7}{c}{\textbf{CWE-125}}\\[-0.8ex]
      \textbf{Same Distribution?}
        & \multicolumn{2}{c|}{No}
        & \multicolumn{2}{c|}{No}
        & \multicolumn{2}{c}{No} \\
      % \textbf{Gaussian distribution}
      %   & No   & No    & Yes & No & Yes & Yes \\
      Mean      & 32.54  & \cellcolor{green!20} \textbf{33.73}  & \textbf{32.81} & \cellcolor{red!20} 31.29 & 35.14 & \cellcolor{green!20} \textbf{42.46} \\
      % Median    & 30.98  & \cellcolor{green!20} \textbf{32.11}  & \textbf{32.88} & \cellcolor{red!20} 30.20 & 35.08 & \cellcolor{green!20} \textbf{42.44} \\
      Std.\ dev.&  \textbf{7.06}  & \cellcolor{green!20} 6.70  & \textbf{8.10} & \cellcolor{green!20} 7.03 & \textbf{7.21} & \cellcolor{green!20} 5.36 \\
      % Skewness  &  0.89  &  1.32  & 0.16 & 0.42 & 0.18 & 0.26 \\
      % Kurtosis  &  1.51  &  2.07  & 0.34 & 1.17 & 0.45 & 0.46 \\
      \bottomrule
    \end{tabular}
  }
  \caption{Comparison of BASE vs GRPO distributions for the top 3 CWE in DiveseVul using the Kolmogorov-Smirnov Test. Bold indicates the highest value; green cells mean GRPO performs better, red worse, and grey highlights distributions that are statistically the same.}
  \label{tab:k-s_sim}
\end{table}
% \begin{figure}[htbp]
%     \centering
%     \begin{subfigure}{0.49\textwidth}
%         \centering
%         \includegraphics[width=\linewidth]{images/New_Diversevul_reasoning_lengths_smoothed_multi_with_means.pdf}
%         \caption{DiverseVul}
%         \label{fig:length_div}
%     \end{subfigure}
%     \hfill
%     \begin{subfigure}{0.49\textwidth}
%         \centering
%         \includegraphics[width=\linewidth]{images/New_BigVul_reasoning_lengths_smoothed_multi_with_means.pdf}
%         \caption{BigVul}
%         \label{fig:fn_beta_small}
%     \end{subfigure}
%     \caption{Answer length comparison between GRPO and SFT for DiverseVul and BigVul}
%     \label{fig:length_big}
% \end{figure}

% Figura 1 – DiverseVul

\subsubsection{Impact of KL Regularization on GRPO}
To better understand the influence of KL regularization on model behaviour, we perform a controlled ablation in which we vary the parameter $\beta$ that regulates the KL divergence weight in GRPO loss. In particular, we compare two values, $\beta=10^{-4}$ and $\beta=10^{-6}$, and observe their effects on the rates of \textbf{True Negatives} (TN) and \textbf{False Negatives} (FN) during training.
\begin{figure}[htbp]
    \centering
        \includegraphics[width=\linewidth]{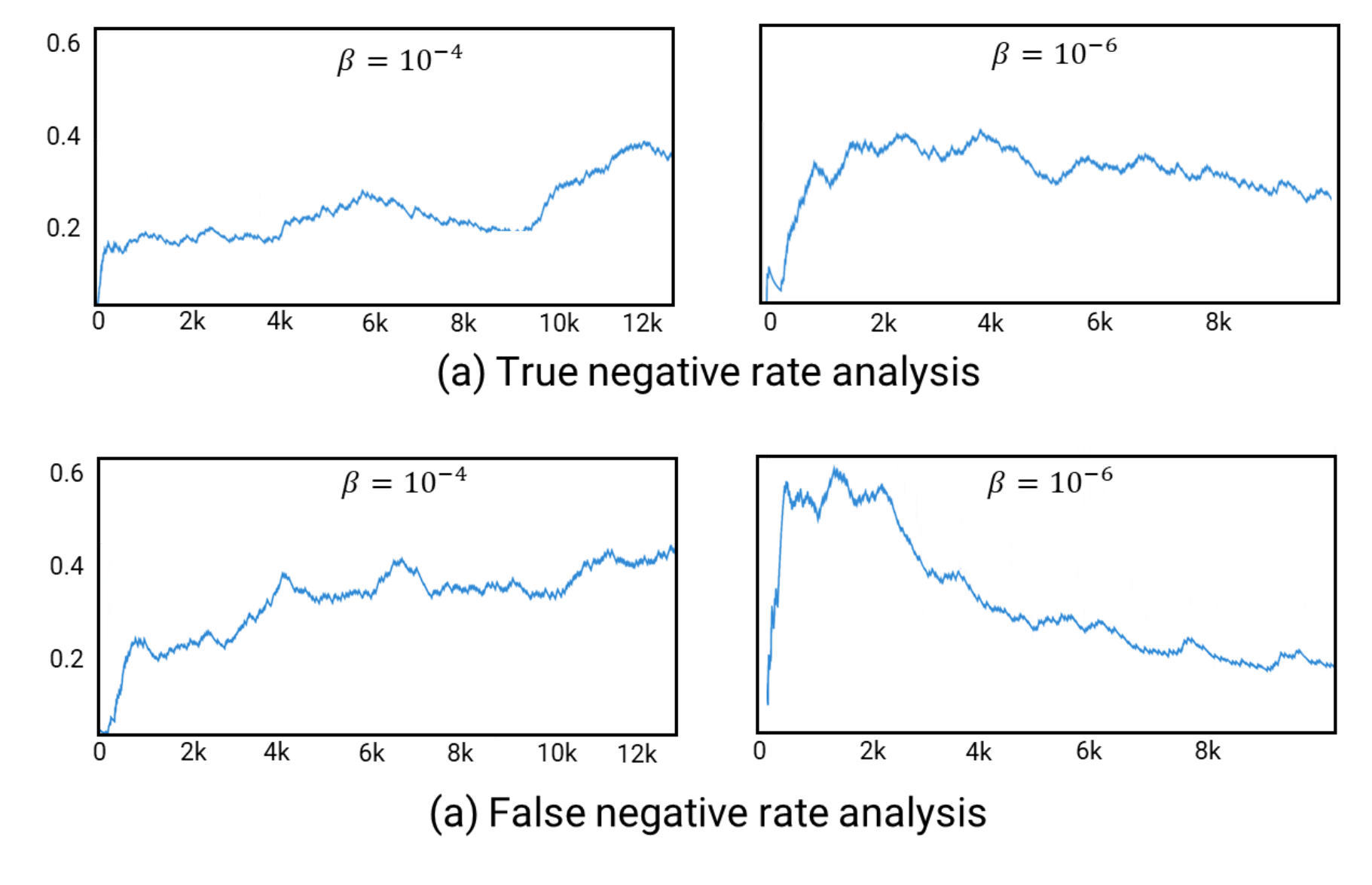}
     \caption{Effect of $\beta$ on \textbf{True Negative} and \textbf{False Negative} rates during training.}
    \label{fig:fn}
\end{figure}
Figure~\ref{fig:fn} illustrates that a higher $\beta$ value leads to a more conservative model: the True Negative rate increases steadily over time, indicating that the model becomes more confident in identifying non-vulnerable code. However, this behaviour comes at a cost: as can be seen in Figure~\ref{fig:fn}, the False Negative rate also increases, especially after the early training phase. This indicates that excessive regularization suppresses risk-taking, which leads to the model under-predicting vulnerabilities in borderline or ambiguous samples.

Conversely, reducing $\beta$ to $10^{-6}$ leads to the opposite result: while the model becomes less confident in rejecting benign samples (lower TN rate), it maintains a lower FN rate throughout training, demonstrating a more risk-taking attitude that prioritises detecting vulnerabilities over avoiding false alarms.

\section{Related Work and Discussion}
\label{sec:related}
Across the literature, various techniques have been proposed to automate vulnerability detection. Prior work can be broadly categorized into three families: \textit{classical deep learning models}, \textit{LLM centric detectors}, and \textit{ behavior oriented LLM prompting}. An overview of these contributions is presented in Table~\ref{tab:diversevul_grpo}, where we distinguish between DNN based and LLM based approaches. 

\paragraph{Classical DNN based approaches} 
Early deep learning methods for vulnerability detection typically embed source code as graphs or token sequences, and then train end to end classifiers using supervised learning. For example, \textit{DeepVulSeeker}~\cite{DeepVulSeeker} combines code graph structures and semantic features, achieving higher accuracy on standard CWE datasets. \textit{VUDENC}~\cite{VUDENC} focuses on real world Python projects, using word2vec embeddings with an LSTM model. \textit{DetectVul}~\cite{DetectVul} removes the need for expensive graph construction by using a self-attention encoder at the statement level. 
Other approaches refine how code structure is modeled. For instance, \textit{TrVD}~\cite{TrVD} breaks the abstract syntax tree (AST) into smaller sub-trees and processes them with a Transformer Lastly, \textit{VulDeeLocator}~\cite{VulDeeLocator} combines token semantics with low level code features to identify vulnerabilities more precisely.

% \paragraph{LLM-centric detectors} Scaling model capacity unlocks stronger semantic priors but raises new challenges. Converting code to LLVM IR, Mahyari~\cite{Harnessing_LLMs} shows that even generic transformers (BERT, DistilBERT) fine-tuned on the IR can match the accuracy of graph networks while remaining language-agnostic. Outside strictly supervised settings, \cite{Analysing_LLM_Capabilities} report that GPT-3.5 reaches parity with specialised detectors and GPT-4 surpasses them, yet both models exhibit brittle behaviour on memory-safety bugs diluted by project-specific idioms.

% Subsequent studies explore how to tailor LLMs to the domain. \textit{Outside the Comfort Zone}~\cite{outside-confort} analyzes large language models’ performance in detecting software vulnerabilities, showing fine-tuning improves smaller models’
% accuracy, highlights dataset mislabeling issues, and
% suggests directions for better training and curation. 

\paragraph{LLM centric detectors}
Recent efforts to automate vulnerability detection have shifted from traditional DNN based pipelines to systems that leverage the reasoning capabilities of LLMs, often by specializing them through ad hoc SFT. For example, Mahyari et al.~\cite{Harnessing_LLMs} demonstrate that even general purpose transformers like BERT and DistilBERT, when finetuned on source code, can match the accuracy of earlier graph based neural networks.
Beyond fully supervised scenarios, \textit{Outside the Comfort Zone} \cite{Analysing_LLM_Capabilities} report that GPT 3.5 performs comparably to specialized vulnerability detectors, while GPT 4 surpasses them. However, both models show fragile behavior when analyzing memory safety bugs, particularly those obscured by project specific coding conventions.
More recent studies explore how to better adapt LLMs to the domain of software vulnerability detection. For example, Guo et al. \cite{Analysing_LLM_Capabilities} and Yang et al. \cite{LLMAO} investigates how LLMs handle vulnerability detection across diverse codebases. 
%The study shows that fine-tuning significantly boosts the performance of smaller models, exposes common dataset mislabeling issues, and offers insights into improving training and data curation practices.

Finally, retrieval augmented generation (RAG) takes LLM based detection a step further. \textit{Vul-RAG}~\cite{VulRAG} builds a knowledge base from CWE descriptions, retrieves relevant facts for each query, and feeds them to the LLM, resulting in higher recall and more interpretable, human readable explanations.

% In another line of work, \textit{LLMAO}~\cite{LLMAO} shows that adapter-based fine-tuning on as few as 100 functions allows a 16B-parameter model to outperform state-of-the-art fault localizers, even without relying on test coverage.

\paragraph{Behaviour oriented LLM prompting} When annotated data are scarce, prompt engineering offers a lightweight alternative, than finetuning approach for LLM, \textit{Prompt-Enhanced ChatGPT} augments zero-shot queries with structural cues and multi turn dialogues~\cite{prompt-chatgpt}. \textit{ProRLearn} treats prompt selection as a RL problem: a policy network iteratively mutates the prompt and receives a reward from the LLM’s output, achieving up to $5\%$ $F_{1}$ improvement over static prompts across three datasets~\cite{ProRLearn}. 
%Such RL approaches avoid catastrophic forgetting but still optimise a black-box objective detached from downstream behaviour.

\paragraph{This work}
In this paper, we take a different approach by enabling the use of a recent GRPO finetuning strategy for LLMs in the context of vulnerability detection, focusing specifically on small, general purpose models. To do so, we introduce a custom multi stage reward function that encourages accurate predictions supported by coherent reasoning, while explicitly penalizing over prediction.
As demonstrated in our extensive experiments, the proposed approach consistently outperforms SFT, highlighting a clear gap with prior work and offering promising directions for future research. It also surpasses prompting based strategies, both those that request reasoning and those that elicit direct answers, showing that GRPO based alignment improves not only decision quality and explanation clarity, but also robustness under domain and language shifts.

\begin{table*}[htbp]
\centering
\footnotesize
\caption{Overview of related work in vulnerability detection. Abbreviations: SL (supervised learning), FT (finetuning), PT (pre-training), PE (prompt engineering), RL (reinforcement learning), IC (in-context/zero-shot).}
\label{tab:diversevul_grpo}
\rowcolors{3}{gray!10}{white}

\begin{adjustbox}{max width=\linewidth}
\setlength{\tabcolsep}{4pt}
\renewcommand{\arraystretch}{1.25}
\begin{tabularx}{\linewidth}{>{\bfseries}l X X >{\centering\arraybackslash}m{1.9cm}}
\toprule
Name (year) & Dataset(s) & Model(s) & Scheme \\
\midrule
\rowcolor{cyan!25}\multicolumn{4}{l}{\textit{Classical DNN-based approaches}}\\
Wang et al.\ (2022)\cite{DeepVulSeeker}
  & Devign\cite{FFmpeg+QEMU},\;SARD\cite{SARD}
  & DeepVulSeeker\cite{DeepVulSeeker}
  & SL \\

Wartschinski et al.\ (2022)\cite{VUDENC}
  & VUDENC\cite{VUDENC}
  & VUDENC\cite{VUDENC}
  & SL \\

Li et al.\ (2022)\cite{VulDeeLocator}
  & NVD\cite{NVD},\;SARD\cite{SARD}
  & BRNN-vdl\cite{VulDeeLocator}
  & SL \\

Tian et al.\ (2024)\cite{TrVD}
  & SARD\cite{SARD}
  & TrVD\cite{TrVD}
  & SL \\

Tran et al.\ (2025)\cite{DetectVul}
  & CVEFixes\cite{CVEFixes},\;VUDENC\cite{VUDENC}
  & DetectVul\cite{DetectVul}
  & SL \\

\midrule
\rowcolor{cyan!25}\multicolumn{4}{l}{\textit{LLM-based approaches}}\\
Mahyari (2024)\cite{Harnessing_LLMs}
  & NVD\cite{NVD},\;SARD\cite{SARD}
  & BERT,\;DistilBERT (cls head)
  & FT \\

Guo et al.\ (2024)\cite{Analysing_LLM_Capabilities}
  & Own data,\;Devign,\;Lin2017,\;Choi2017,\;LineVul,\;PrimeVul
  & VulBERTa,\;CodeBERT,\;Mistral,\;Mixtral 8×7B,\;CodeLLaMA,\;GPT-4
  & IC \\

Yang et al.\ (2024)\cite{LLMAO}
  & Defects4J\cite{Defects4J},\;Devign\cite{FFmpeg+QEMU},\;BugsInPy\cite{BugsInPy}
  & LLMAO\cite{LLMAO},\;CodeGen\cite{CodeGen}
  & PT\,+\,FT \\

Zhang et al.\ (2024)\cite{prompt-chatgpt}
  & SARD\cite{SARD},\;SySeVR\cite{SySeVR}
  & ChatGPT-4\cite{GPT4}
  & PE \\

Ren et al.\ (2024)\cite{ProRLearn}
  & Devign\cite{FFmpeg+QEMU},\;Reveal\cite{reveal},\;BigVul\cite{bigvul}
  & CodeBERT\cite{CodeBERT}
  & RL \\
\bottomrule
\end{tabularx}
\end{adjustbox}
\end{table*}

\section{Conclusion}
\label{sec:conclusion} 

\begin{table*}[!htbp]
  \centering
  \small            % carattere leggermente più piccolo
  \caption{Percentage improvement (\textcolor{green!50!black}{+}) or deterioration
           (\textcolor{red!60!black}{--}) of GRPO across all datasets.}
  \label{tab:grpo_overall}

  % ---------- (a) ----------
  \begin{subtable}[t]{\linewidth}
    \centering
    \caption{\textsc{DIVERSEVUL}}
    \resizebox{0.85\linewidth}{!}{%
      % \begin{table*}[htbp]
%     \centering
%     \small
%     \caption{Percentage improvement (\textcolor{green!50!black}{+}) or deterioration (\textcolor{red!60!black}{--}) of GRPO over vs No Reas., Reas. and SFT on \textsc{DIVERSEVUL}}
%     \setlength{\tabcolsep}{4pt}
%     \renewcommand{\arraystretch}{1.1}
%     \rowcolors{3}{gray!8}{white}
%     \resizebox{0.98\linewidth}{!}{%
\begin{tabular}{>{\raggedright\arraybackslash}p{3.5cm}|%
                >{\centering\arraybackslash}p{1.4cm}>{\centering\arraybackslash}p{1.4cm}>{\centering\arraybackslash}p{1.4cm}|%
                >{\centering\arraybackslash}p{1.4cm}>{\centering\arraybackslash}p{1.4cm}>{\centering\arraybackslash}p{1.4cm}|%
                >{\centering\arraybackslash}p{1.4cm}>{\centering\arraybackslash}p{1.4cm}>{\centering\arraybackslash}p{1.4cm}}
    \toprule
    \rowcolor{gray!25}
    \textbf{Metric} &
    \multicolumn{3}{c|}{\textbf{Qwen 2.5}} &
    \multicolumn{3}{c|}{\textbf{LLaMA 3B}} &
    \multicolumn{3}{c}{\textbf{LLaMA 8B}} \\
    \rowcolor{gray!25}
    & \textit{vs No Reas.} & \textit{vs Reas.} & \textit{vs SFT} &
      \textit{vs No Reas.} & \textit{vs Reas.} & \textit{vs SFT} &
      \textit{vs No Reas.} & \textit{vs Reas.} & \textit{vs SFT} \\
    \midrule
    \rowcolor{gray!10}\multicolumn{10}{l}{\textbf{Not Vulnerable}} \\
    \textit{Precision}          & \textcolor{green!50!black}{+32.30\%} & \textcolor{green!50!black}{+43.10\%} & \textcolor{green!50!black}{+46.00\%} & \textcolor{red!60!black}{--18.20\%} & \textcolor{green!50!black}{+19.10\%} & \textcolor{red!60!black}{--27.90\%} & \textcolor{green!50!black}{+100\%} & \textcolor{green!50!black}{+31.60\%} & \textcolor{red!60!black}{--34.20\%} \\
    \textit{Recall}             & \textcolor{green!50!black}{+11.60\%} & \textcolor{red!60!black}{--18.10\%} & \textcolor{green!50!black}{+777.10\%} & \textcolor{green!50!black}{+100\%} & \textcolor{green!50!black}{+38.50\%} & \textcolor{green!50!black}{+125.00\%} & \textcolor{green!50!black}{+100\%} & \textcolor{green!50!black}{+62.30\%} & \textcolor{red!60!black}{--1.90\%} \\
    \textit{F\textsubscript{1}} & \textcolor{green!50!black}{+21.30\%} & \textcolor{green!50!black}{+11.00\%} & \textcolor{green!50!black}{+445.80\%} & \textcolor{green!50!black}{+100\%} & \textcolor{green!50!black}{+29.50\%} & \textcolor{green!50!black}{+47.00\%} & \textcolor{green!50!black}{+100\%} & \textcolor{green!50!black}{+59.40\%} & \textcolor{red!60!black}{--9.30\%} \\
    \rowcolor{gray!10}\multicolumn{10}{l}{\textbf{Vulnerable}} \\
    \textit{Precision}          & \textcolor{green!50!black}{+32.30\%} & \textcolor{green!50!black}{+43.10\%} & \textcolor{green!50!black}{+46.00\%} & \textcolor{red!60!black}{--18.20\%} & \textcolor{green!50!black}{+19.10\%} & \textcolor{red!60!black}{--27.90\%} & \textcolor{green!50!black}{+100\%} & \textcolor{green!50!black}{+31.60\%} & \textcolor{red!60!black}{--34.20\%} \\
    \textit{Recall}             & \textcolor{green!50!black}{+11.60\%} & \textcolor{red!60!black}{--18.10\%} & \textcolor{green!50!black}{+777.10\%} & \textcolor{green!50!black}{+100\%} & \textcolor{green!50!black}{+38.50\%} & \textcolor{green!50!black}{+125.00\%} & \textcolor{green!50!black}{+100\%} & \textcolor{green!50!black}{+62.30\%} & \textcolor{red!60!black}{--1.90\%} \\
    \textit{F\textsubscript{1}} & \textcolor{green!50!black}{+21.30\%} & \textcolor{green!50!black}{+11.00\%} & \textcolor{green!50!black}{+445.80\%} & \textcolor{green!50!black}{+100\%} & \textcolor{green!50!black}{+29.50\%} & \textcolor{green!50!black}{+47.00\%} & \textcolor{green!50!black}{+100\%} & \textcolor{green!50!black}{+59.40\%} & \textcolor{red!60!black}{--9.30\%} \\
    \rowcolor{gray!10}\multicolumn{10}{l}{\textbf{Overall}} \\
    \textit{Accuracy}           & \textcolor{green!50!black}{+29.20\%} & \textcolor{green!50!black}{+40.00\%} & \textcolor{green!50!black}{+34.40\%} & \textcolor{green!50!black}{+17.30\%} & \textcolor{green!50!black}{+27.80\%} & \textcolor{green!50!black}{+8.50\%}  & \textcolor{green!50!black}{+25.70\%} & \textcolor{green!50!black}{+36.90\%} & \textcolor{red!60!black}{--0.60\%} \\
    \textit{Macro Precision}    & \textcolor{green!50!black}{+29.80\%} & \textcolor{green!50!black}{+46.70\%} & \textcolor{green!50!black}{+37.80\%} & \textcolor{green!50!black}{+11.90\%} & \textcolor{green!50!black}{+44.20\%} & \textcolor{green!50!black}{+22.50\%} & \textcolor{green!50!black}{+156.00\%} & \textcolor{green!50!black}{+42.20\%} & \textcolor{green!50!black}{+0.00\%} \\
    \textit{Macro Recall}       & \textcolor{green!50!black}{+29.40\%} & \textcolor{green!50!black}{+43.20\%} & \textcolor{green!50!black}{+34.60\%} & \textcolor{green!50!black}{+36.80\%} & \textcolor{green!50!black}{+48.70\%} & \textcolor{green!50!black}{+17.90\%} & \textcolor{green!50!black}{+23.60\%} & \textcolor{red!60!black}{--4.90\%} & \textcolor{red!60!black}{--0.30\%} \\
    \textit{Macro F\textsubscript{1}} & \textcolor{green!50!black}{+29.00\%} & \textcolor{green!50!black}{+56.00\%} & \textcolor{green!50!black}{+76.60\%} & \textcolor{green!50!black}{+61.80\%} & \textcolor{green!50!black}{+18.70\%} & \textcolor{green!50!black}{+2.70\%}  & \textcolor{green!50!black}{+82.40\%} & \textcolor{green!50!black}{+40.00\%} & \textcolor{green!50!black}{+0.30\%} \\
    \textit{Weighted Precision} & \textcolor{green!50!black}{+29.80\%} & \textcolor{green!50!black}{+46.70\%} & \textcolor{green!50!black}{+37.80\%} & \textcolor{green!50!black}{+12.20\%} & \textcolor{green!50!black}{+44.70\%} & \textcolor{green!50!black}{+22.80\%} & \textcolor{green!50!black}{+166.70\%} & \textcolor{green!50!black}{+42.20\%} & \textcolor{green!50!black}{+0.00\%} \\
    \textit{Weighted Recall}    & \textcolor{green!50!black}{+29.20\%} & \textcolor{green!50!black}{+40.00\%} & \textcolor{green!50!black}{+34.40\%} & \textcolor{green!50!black}{+17.30\%} & \textcolor{green!50!black}{+27.80\%} & \textcolor{green!50!black}{+8.50\%}  & \textcolor{green!50!black}{+25.70\%} & \textcolor{green!50!black}{+36.90\%} & \textcolor{red!60!black}{--0.60\%} \\
    \textit{Weighted F\textsubscript{1}} & \textcolor{green!50!black}{+29.00\%} & \textcolor{green!50!black}{+56.00\%} & \textcolor{green!50!black}{+76.60\%} & \textcolor{green!50!black}{+61.80\%} & \textcolor{green!50!black}{+18.70\%} & \textcolor{green!50!black}{+2.70\%}  & \textcolor{green!50!black}{+82.10\%} & \textcolor{green!50!black}{+39.80\%} & \textcolor{green!50!black}{+0.20\%} \\
    \bottomrule
\end{tabular}

    % }
%     \label{tab:grpo_improvement_diversevul}
% \end{table*} % oppure incolla qui il tabular
    }
  \end{subtable}

  \medskip  % piccolo spazio verticale (si può ridurre a \smallskip)

  % ---------- (b) ----------
  \begin{subtable}[t]{\linewidth}
    \centering
    \caption{\textsc{CLEANVUL}}
    \resizebox{0.85\linewidth}{!}{%
\begin{tabular}{>{\raggedright\arraybackslash}p{3.5cm}|%
                >{\centering\arraybackslash}p{1.4cm}>{\centering\arraybackslash}p{1.4cm}>{\centering\arraybackslash}p{1.4cm}|%
                >{\centering\arraybackslash}p{1.4cm}>{\centering\arraybackslash}p{1.4cm}>{\centering\arraybackslash}p{1.4cm}|%
                >{\centering\arraybackslash}p{1.4cm}>{\centering\arraybackslash}p{1.4cm}>{\centering\arraybackslash}p{1.4cm}}
    \toprule
    \rowcolor{gray!25}
    \textbf{Metric} &
    \multicolumn{3}{c|}{\textbf{Qwen 2.5}} &
    \multicolumn{3}{c|}{\textbf{LLaMA 3B}} &
    \multicolumn{3}{c}{\textbf{LLaMA 8B}} \\
    \rowcolor{gray!25}
    & \textit{vs No Reas.} & \textit{vs Reas.} & \textit{vs SFT} &
      \textit{vs No Reas.} & \textit{vs Reas.} & \textit{vs SFT} &
      \textit{vs No Reas.} & \textit{vs Reas.} & \textit{vs SFT} \\
    \midrule
    \rowcolor{gray!10}\multicolumn{10}{l}{\textbf{Not Vulnerable}} \\
    \textit{Precision}          & \textcolor{green!50!black}{+2.90\%} & \textcolor{green!50!black}{+10.8\%} & \textcolor{green!50!black}{+28.00\%} & \textcolor{red!60!black}{--42.10\%} & \textcolor{green!50!black}{+9.2\%} & \textcolor{green!50!black}{+5.30\%} & \textcolor{red!60!black}{--30.00\%} & \textcolor{green!50!black}{+40.0\%} & \textcolor{green!50!black}{+34.60\%} \\
    \textit{Recall}             & \textcolor{red!60!black}{--16.60\%} & \textcolor{red!60!black}{--32.2\%} & \textcolor{green!50!black}{+577.50\%} & \textcolor{green!50!black}{+100\%} & \textcolor{green!50!black}{+46.2\%} & \textcolor{green!50!black}{+587.00\%} & \textcolor{green!50!black}{+100\%} & \textcolor{red!60!black}{--53.9\%} & \textcolor{green!50!black}{+371.10\%} \\
    \textit{F1}                 & \textcolor{red!60!black}{--6.80\%} & \textcolor{red!60!black}{--11.3\%} & \textcolor{green!50!black}{+330.00\%} & \textcolor{green!50!black}{+100\%} & \textcolor{green!50!black}{+25.6\%} & \textcolor{green!50!black}{+269.40\%} & \textcolor{green!50!black}{+100\%} & \textcolor{red!60!black}{--23.1\%} & \textcolor{green!50!black}{+212.50\%} \\
    \rowcolor{gray!10}\multicolumn{10}{l}{\textbf{Vulnerable}} \\
    \textit{Precision}          & \textcolor{green!50!black}{+2.90\%} & \textcolor{green!50!black}{+1.4\%}  & \textcolor{green!50!black}{+28.00\%} & \textcolor{red!60!black}{--42.10\%} & \textcolor{green!50!black}{+19.0\%} & \textcolor{green!50!black}{+5.30\%} & \textcolor{red!60!black}{--30.00\%} & \textcolor{green!50!black}{+5.5\%}  & \textcolor{green!50!black}{+34.60\%} \\
    \textit{Recall}             & \textcolor{red!60!black}{--16.60\%} & \textcolor{green!50!black}{+131.2\%} & \textcolor{green!50!black}{+577.50\%} & \textcolor{green!50!black}{+100\%} & \textcolor{red!60!black}{--12.9\%} & \textcolor{green!50!black}{+587.00\%} & \textcolor{green!50!black}{+100\%} & \textcolor{green!50!black}{+710.0\%} & \textcolor{green!50!black}{+371.10\%} \\
    \textit{F1}                 & \textcolor{red!60!black}{--6.80\%} & \textcolor{green!50!black}{+66.9\%} & \textcolor{green!50!black}{+330.00\%} & \textcolor{green!50!black}{+100\%} & \textcolor{green!50!black}{+1.1\%} & \textcolor{green!50!black}{+269.40\%} & \textcolor{green!50!black}{+100\%} & \textcolor{green!50!black}{+322.0\%} & \textcolor{green!50!black}{+212.50\%} \\
    \rowcolor{gray!10}\multicolumn{10}{l}{\textbf{Overall}} \\
    \textit{Accuracy}           & \textcolor{green!50!black}{+0.40\%} & \textcolor{green!50!black}{+7.9\%}  & \textcolor{green!50!black}{+16.70\%} & \textcolor{green!50!black}{+19.00\%} & \textcolor{green!50!black}{+12.3\%} & \textcolor{green!50!black}{+16.70\%} & \textcolor{green!50!black}{+15.40\%} & \textcolor{green!50!black}{+15.4\%} & \textcolor{green!50!black}{+15.40\%} \\
    \textit{Macro Precision}    & \textcolor{green!50!black}{+0.40\%} & \textcolor{green!50!black}{+5.9\%}  & \textcolor{green!50!black}{+21.70\%} & \textcolor{red!60!black}{--20.10\%} & \textcolor{green!50!black}{+13.0\%} & \textcolor{green!50!black}{+13.00\%} & \textcolor{red!60!black}{--22.10\%} & \textcolor{green!50!black}{+12.3\%} & \textcolor{green!50!black}{+14.50\%} \\
    \textit{Macro Recall}       & \textcolor{green!50!black}{+0.40\%} & \textcolor{green!50!black}{+7.9\%}  & \textcolor{green!50!black}{+16.70\%} & \textcolor{green!50!black}{+19.20\%} & \textcolor{green!50!black}{+12.5\%} & \textcolor{green!50!black}{+16.90\%} & \textcolor{green!50!black}{+15.40\%} & \textcolor{green!50!black}{+15.4\%} & \textcolor{green!50!black}{+15.40\%} \\
    \textit{Macro F\textsubscript{1}} & \textcolor{green!50!black}{+0.20\%} & \textcolor{green!50!black}{+16.5\%} & \textcolor{green!50!black}{+46.40\%} & \textcolor{green!50!black}{+79.40\%} & \textcolor{green!50!black}{+11.7\%} & \textcolor{green!50!black}{+44.40\%} & \textcolor{green!50!black}{+71.80\%} & \textcolor{green!50!black}{+41.7\%} & \textcolor{green!50!black}{+41.70\%} \\
    \textit{Weighted Precision} & \textcolor{green!50!black}{+0.40\%} & \textcolor{green!50!black}{+5.9\%}  & \textcolor{green!50!black}{+21.70\%} & \textcolor{red!60!black}{--20.10\%} & \textcolor{green!50!black}{+13.0\%} & \textcolor{green!50!black}{+13.00\%} & \textcolor{red!60!black}{--22.10\%} & \textcolor{green!50!black}{+12.3\%} & \textcolor{green!50!black}{+14.50\%} \\
    \textit{Weighted Recall}    & \textcolor{green!50!black}{+0.40\%} & \textcolor{green!50!black}{+7.9\%}  & \textcolor{green!50!black}{+16.70\%} & \textcolor{green!50!black}{+19.00\%} & \textcolor{green!50!black}{+12.3\%} & \textcolor{green!50!black}{+16.70\%} & \textcolor{green!50!black}{+15.40\%} & \textcolor{green!50!black}{+15.4\%} & \textcolor{green!50!black}{+15.40\%} \\
    \textit{Weighted F\textsubscript{1}} & \textcolor{green!50!black}{+0.20\%} & \textcolor{green!50!black}{+16.5\%} & \textcolor{green!50!black}{+46.40\%} & \textcolor{green!50!black}{+65.50\%} & \textcolor{green!50!black}{+11.7\%} & \textcolor{green!50!black}{+44.40\%} & \textcolor{green!50!black}{+88.50\%} & \textcolor{green!50!black}{+41.7\%} & \textcolor{green!50!black}{+41.70\%} \\
    \bottomrule
\end{tabular}

    }
  \end{subtable}

  \medskip

  % ---------- (c) ----------
  \begin{subtable}[t]{\linewidth}
    \centering
    \caption{\textsc{BIGVUL}}
    \resizebox{0.85\linewidth}{!}{%

\begin{tabular}{>{\raggedright\arraybackslash}p{3.5cm}|%
                >{\centering\arraybackslash}p{1.4cm}>{\centering\arraybackslash}p{1.4cm}>{\centering\arraybackslash}p{1.4cm}|%
                >{\centering\arraybackslash}p{1.4cm}>{\centering\arraybackslash}p{1.4cm}>{\centering\arraybackslash}p{1.4cm}|%
                >{\centering\arraybackslash}p{1.4cm}>{\centering\arraybackslash}p{1.4cm}>{\centering\arraybackslash}p{1.4cm}}
    \toprule
    \rowcolor{gray!25}
    \textbf{Metric} &
    \multicolumn{3}{c|}{\textbf{Qwen 2.5}} &
    \multicolumn{3}{c|}{\textbf{LLaMA 3B}} &
    \multicolumn{3}{c}{\textbf{LLaMA 8B}} \\
    \rowcolor{gray!25}
    & \textit{ vs No Reas.} & \textit{vs Reas.} & \textit{vs SFT} &
      \textit{ vs No Reas.} & \textit{vs Reas.} & \textit{vs SFT} &
      \textit{ vs No Reas.} & \textit{vs Reas.} & \textit{vs SFT} \\
    \midrule
    \rowcolor{gray!10}\multicolumn{10}{l}{\textbf{Not Vulnerable}} \\
    \textit{Precision}          & \textcolor{green!50!black}{+13.0\%} & \textcolor{green!50!black}{+17.2\%} & \textcolor{green!50!black}{+15.10\%} & \textcolor{green!50!black}{+100\%} & \textcolor{green!50!black}{+6.7\%} & \textcolor{green!50!black}{+0.90\%}  & \textcolor{red!60!black}{--36.3\%} & \textcolor{green!50!black}{+24.9\%} & \textcolor{green!50!black}{+4.40\%} \\
    \textit{Recall}             & \textcolor{red!60!black}{--24.5\%} & \textcolor{red!60!black}{--21.0\%} & \textcolor{green!50!black}{+428.50\%} & \textcolor{green!50!black}{+100\%} & \textcolor{green!50!black}{+21.3\%} & \textcolor{green!50!black}{+394.60\%} & \textcolor{green!50!black}{+100\%} & \textcolor{red!60!black}{--38.6\%} & \textcolor{green!50!black}{+320.70\%} \\
    \textit{F\textsubscript{1}} & \textcolor{red!60!black}{--4.5\%}  & \textcolor{red!60!black}{--0.2\%}  & \textcolor{green!50!black}{+199.50\%} & \textcolor{green!50!black}{+100\%} & \textcolor{green!50!black}{+14.0\%} & \textcolor{green!50!black}{+182.40\%} & \textcolor{green!50!black}{+100\%} & \textcolor{red!60!black}{--8.7\%}  & \textcolor{green!50!black}{+166.10\%} \\
    \rowcolor{gray!10}\multicolumn{10}{l}{\textbf{Vulnerable}} \\
    \textit{Precision}          & \textcolor{green!50!black}{+13.0\%} & \textcolor{green!50!black}{+17.2\%} & \textcolor{green!50!black}{+15.10\%} & \textcolor{green!50!black}{+100\%} & \textcolor{green!50!black}{+6.7\%} & \textcolor{green!50!black}{+0.90\%}  & \textcolor{red!60!black}{--36.3\%} & \textcolor{green!50!black}{+24.9\%} & \textcolor{green!50!black}{+4.40\%} \\
    \textit{Recall}             & \textcolor{red!60!black}{--24.5\%} & \textcolor{red!60!black}{--21.0\%} & \textcolor{green!50!black}{+428.50\%} & \textcolor{green!50!black}{+100\%} & \textcolor{green!50!black}{+21.3\%} & \textcolor{green!50!black}{+394.60\%} & \textcolor{green!50!black}{+100\%} & \textcolor{red!60!black}{--38.6\%} & \textcolor{green!50!black}{+320.70\%} \\
    \textit{F\textsubscript{1}} & \textcolor{red!60!black}{--4.5\%}  & \textcolor{red!60!black}{--0.2\%}  & \textcolor{green!50!black}{+199.50\%} & \textcolor{green!50!black}{+100\%} & \textcolor{green!50!black}{+14.0\%} & \textcolor{green!50!black}{+182.40\%} & \textcolor{green!50!black}{+100\%} & \textcolor{red!60!black}{--8.7\%}  & \textcolor{green!50!black}{+166.10\%} \\
    \rowcolor{gray!10}\multicolumn{10}{l}{\textbf{Overall}} \\
    \textit{Accuracy}           & \textcolor{green!50!black}{+8.5\%}  & \textcolor{green!50!black}{+16.4\%} & \textcolor{green!50!black}{+25.50\%} & \textcolor{green!50!black}{+12.2\%} & \textcolor{green!50!black}{+7.8\%} & \textcolor{green!50!black}{+7.80\%}  & \textcolor{green!50!black}{+26.9\%} & \textcolor{green!50!black}{+22.0\%} & \textcolor{green!50!black}{+19.60\%} \\
    \textit{Macro Precision}    & \textcolor{red!60!black}{--1.5\%}  & \textcolor{green!50!black}{+10.3\%} & \textcolor{green!50!black}{+20.80\%} & \textcolor{green!50!black}{+120.0\%} & \textcolor{green!50!black}{+7.8\%} & \textcolor{green!50!black}{+3.80\%}  & \textcolor{red!60!black}{--16.9\%} & \textcolor{green!50!black}{+17.5\%} & \textcolor{green!50!black}{+11.20\%} \\
    \textit{Macro Recall}       & \textcolor{green!50!black}{+8.3\%}  & \textcolor{green!50!black}{+16.2\%} & \textcolor{green!50!black}{+25.30\%} & \textcolor{green!50!black}{+9.6\%}  & \textcolor{green!50!black}{+7.5\%} & \textcolor{green!50!black}{+7.50\%}  & \textcolor{green!50!black}{+24.6\%} & \textcolor{green!50!black}{+22.2\%} & \textcolor{green!50!black}{+17.50\%} \\
    \textit{Macro F\textsubscript{1}} & \textcolor{green!50!black}{+18.1\%} & \textcolor{green!50!black}{+27.6\%} & \textcolor{green!50!black}{+48.40\%} & \textcolor{green!50!black}{+65.2\%} & \textcolor{green!50!black}{+6.9\%} & \textcolor{green!50!black}{+29.80\%} & \textcolor{green!50!black}{+88.5\%} & \textcolor{green!50!black}{+63.7\%} & \textcolor{green!50!black}{+41.40\%} \\
    \textit{Weighted Precision} & \textcolor{red!60!black}{--1.5\%}  & \textcolor{green!50!black}{+10.3\%} & \textcolor{green!50!black}{+20.80\%} & \textcolor{green!50!black}{+129.2\%} & \textcolor{green!50!black}{+7.8\%} & \textcolor{green!50!black}{+3.80\%}  & \textcolor{red!60!black}{--16.9\%} & \textcolor{green!50!black}{+17.5\%} & \textcolor{green!50!black}{+11.20\%} \\
    \textit{Weighted Recall}    & \textcolor{green!50!black}{+8.5\%}  & \textcolor{green!50!black}{+16.4\%} & \textcolor{green!50!black}{+25.50\%} & \textcolor{green!50!black}{+12.2\%} & \textcolor{green!50!black}{+7.8\%} & \textcolor{green!50!black}{+7.80\%}  & \textcolor{green!50!black}{+26.9\%} & \textcolor{green!50!black}{+22.0\%} & \textcolor{green!50!black}{+19.60\%} \\
    \textit{Weighted F\textsubscript{1}} & \textcolor{green!50!black}{+16.2\%} & \textcolor{green!50!black}{+27.8\%} & \textcolor{green!50!black}{+48.60\%} & \textcolor{green!50!black}{+65.5\%} & \textcolor{green!50!black}{+7.1\%} & \textcolor{green!50!black}{+30.00\%} & \textcolor{green!50!black}{+88.5\%} & \textcolor{green!50!black}{+59.5\%} & \textcolor{green!50!black}{+41.40\%} \\
    \bottomrule
\end{tabular}

    }
  \end{subtable}
    \label{tab:all}
\end{table*}

This paper revisits vulnerability detection with LLMs from the perspective of \emph{reasoning aware} alignment, formulating the study around three research questions (RQ1–RQ3). 

\RQ{RQ1} raised the question of whether commercially available instruction tuned models can reliably reason about code. Experimental results suggest that these models struggle to fully leverage their reasoning capabilities. Even in the case of LLaMA, where reasoning aloud leads to some improvements, predictions remain unbalanced and prone to false negatives (e.g. LLaMA 8b).
Building on these findings, \RQ{RQ2} asked whether a model can be finetuned to better exploit its own reasoning ability. To address this, we enabled the use of GRPO by designing a dynamic reward function that balances formatting, reasoning, and correctness, while adapting over time to prevent \textit{reward hacking}.
%A single epoch of GRPO, guided by this reward, increases recall for the \emph{Not Vulnerable} class from 0.07 to 0.61 without losing precision, and achieves similar or better gains on out-of-distribution datasets. 

Finally, \RQ{RQ3} addresses a comparison between GRPO and SFT across all the three benchmarks. The extensive set of results shows that GRPO outperforms SFT by 1–29 macro F$_1$ points and improves accuracy by 4–17 points.
When focusing on the top 10 CWEs from the 2024 CWE Top 25 list, GRPO demonstrates consistent improvements across most vulnerability types and model sizes.
Moreover, GRPO generalizes better than SFT across diverse vulnerability types, programming languages, and data distributions.
Multiple ablation studies have also been conducted, highlighting that GRPO not only improves detection accuracy but also enhances the quality of the model reasoning.

To conclude, the proposed formulation of GRPO for vulnerability detection enables LLMs to reason more effectively about software flaws, resulting in decisions that are not only more accurate but also more interpretable.
By aligning model outputs with security specific reasoning patterns, without relying on explicitly supervised rationales during training, the presented work takes a step toward trustworthy and scalable vulnerability analysis with generative models. 

% Macro-F\textsubscript{1} increases steadily on Java, Python, and JavaScript, with the largest relative gains on Python, while accuracy and weighted-F\textsubscript{1} follow the same trend across all models.

% Ablation studies show that GRPO not only improves detection accuracy, but also leads to better reasoning.  
% Explanations align more closely with CWE definitions, become 40–50\% shorter without losing correctness, and are more focused overall.  
% A smaller KL regularization term ($\beta = 10^{-6}$) further improves the balance between true and false negatives, while larger values make the model too conservative.  
% In sum, GRPO promotes clearer, more accurate, and security-aware reasoning than standard finetuning. Table \ref{tab:all} shows the percentages of improvement or deterioration of GRPO across all datasets with respect to SFT and Prompt engineering (\textit{Reasoning} and \textit{No Reasoning}).

In summary, GRPO enables small language models to reason more effectively about software vulnerabilities, producing decisions that are not only more accurate but also more interpretable.  
By aligning model outputs with security specific reasoning patterns, without relying on explicit supervised rationales during training, our approach takes a step toward trustworthy and scalable vulnerability analysis with foundation models.

    \bibliographystyle{elsarticle-num}
    \bibliography{main}
    \clearpage
    % \appendix
    % \input{appendix}
\end{document}